\documentclass[aip,jcp,reprint,amsmath,amssymb]{revtex4-2}
\usepackage{graphicx}
\usepackage{natmove}

\begin{document}

\title{Discretization strategies for colloidal particles in multiparticle collision dynamics simulations}

\author{Michaela Bush}
\affiliation{Department of Chemical Engineering, Auburn University, Auburn, AL 36849, USA}

\author{Michael P. Howard}
\email{mphoward@auburn.edu}
\affiliation{Department of Chemical Engineering, Auburn University, Auburn, AL 36849, USA}

\begin{abstract}
Discrete models are frequently used in multiparticle dynamics simulations to capture hydrodynamic interactions between colloidal particles and the solvent as well as to represent anisotropic pairwise interactions between colloidal particles; however, there is currently limited guidance on how to reliably parameterize these models. Here, we first compare strategies for selecting the density and mass of discrete surface sites used to couple colloidal particles to the solvent, finding that using a minimum of 2 sites per unit area with a scheme that matches the total mass and moment of inertia for a neutrally buoyant solid particle produces reliable and accurate results for the transport properties of colloidal particles at both infinite dilution and in suspension. We then compare strategies for representing the excluded volume of nearly-hard shape-anisotropic colloidal particles using a collection of discrete interaction sites with isotropic repulsion, finding that having a discrepancy between the nominal volume and the excluded volume of the particle can significantly affect suspension transport properties. This discrepancy can be mitigated by placing the interaction sites inside and tangent to the surface of the colloidal particle. We expect these findings to help construct discrete models for colloidal particles with less sensitivity to parameterization.
\end{abstract}

\maketitle

\section{Introduction}
The dynamics of colloidal particles suspended in a solvent play an important role in numerous processes, ranging from the drying-induced assembly of nanoparticles \cite{boles_self-assembly_2016, howard_evaporation-induced_2018, kundu_exploring_2025} to the heteroaggregation of microplastics \cite{rusenargun_influence_2023}. Computer simulations can provide insights needed to understand and engineer these processes, but simulating colloidal suspensions is challenging because the length and time scales associated with the solvent are typically orders of magnitude smaller than those associated with the colloidal particles, making it frequently impractical to fully resolve both \cite{padding_hydrodynamic_2006, howard_modeling_2019}. Several methods have been developed to reduce the computational cost of these simulations by using other means to capture the effects of the solvent on the colloidal particles \cite{brady_stokesian_1988, chen_lattice_1998, liu_dissipative_2015}. One such method is multiparticle collision dynamics (MPCD) \cite{malevanets_mesoscopic_1999, kapral_multiparticle_2008, gompper_multi-particle_2009, howard_modeling_2019}, which is the focus of this article.

\begin{figure}[hb]
    \centering
    \includegraphics{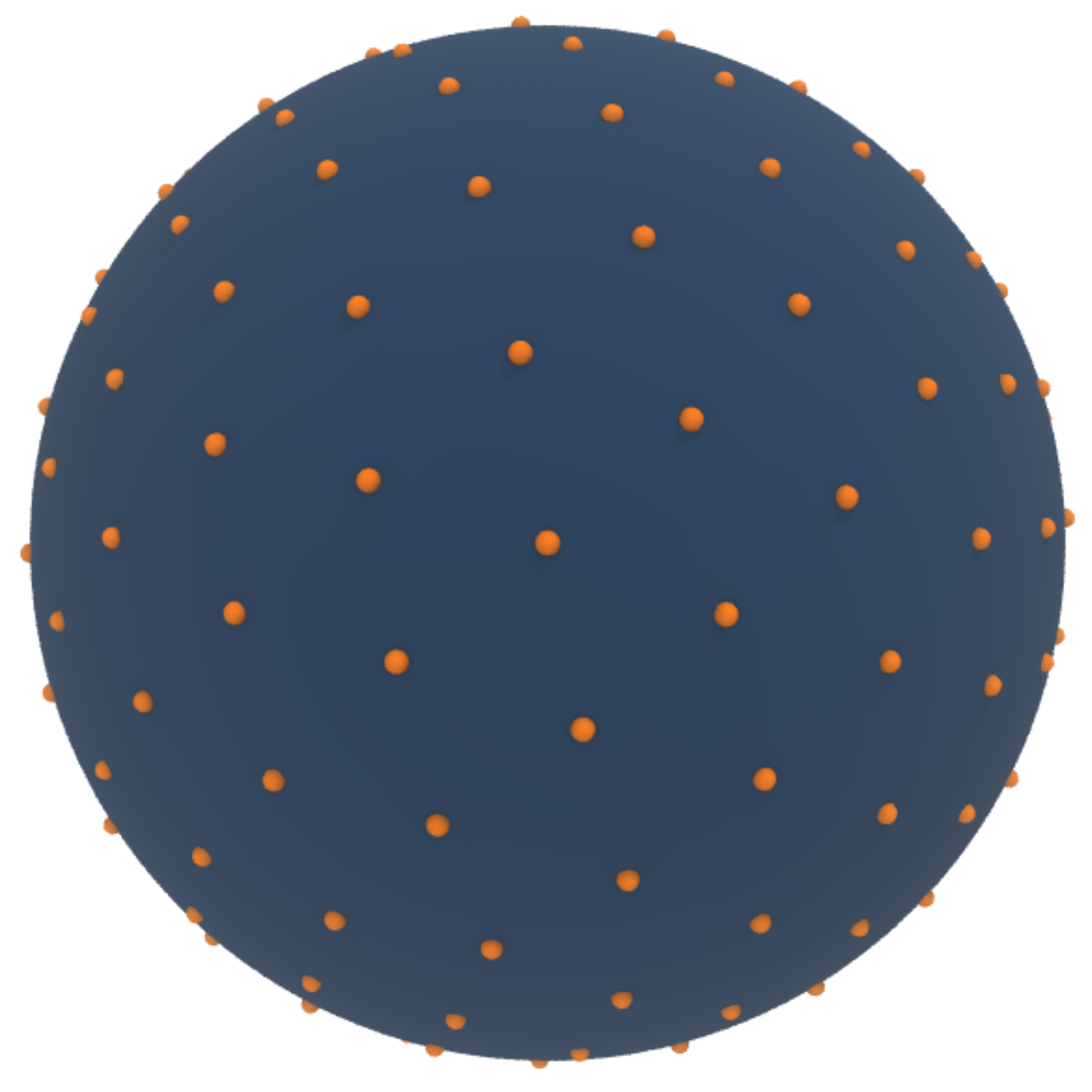}
    \caption{Sphere with diameter $d=6\,\ell$ (blue) and discrete surface sites (orange) at density $\alpha = 2\,\ell^{-2}$. Image rendered using VMD 1.9.3 \cite{HUMP96}.}
    \label{fig:discrete_particle}
\end{figure}

MPCD is a mesoscale particle-based simulation method that uses an explicit but highly simplified solvent model. The MPCD solvent consists of point particles that do not interact through forces; instead, they exchange momentum with each other through stochastic collisions in spatially localized cells \cite{malevanets_mesoscopic_1999}. Colloidal particles must be coupled to the solvent in a way that leads to the build up of solvent-mediated hydrodynamic interactions. Numerous coupling strategies have been proposed \cite{malevanets_dynamics_2000, padding_hydrodynamic_2006, padding_stick_2005, whitmer_fluidsolid_2010}, but the current best practice is to discretize the surface of the colloidal particle into sites (Fig.~\ref{fig:discrete_particle}) that also participate in the solvent's momentum-exchanging collisions \cite{poblete_hydrodynamics_2014, wani_diffusion_2022, wani_mesoscale_2024, peng_multiparticle_2024, howard_transport_2026-1, bush_simulating_2026}. Between collisions, both the solvent and the colloidal particles evolve according to Newton's equations of motion, with the interactions between colloidal particles represented by a potential energy function.

MPCD simulations using discrete particle models have been shown to be effective at modeling many transport properties of hard-sphere colloidal particles \cite{wani_diffusion_2022,peng_multiparticle_2024,howard_transport_2026-1,bush_simulating_2026}; however, it has been observed that the results can be sensitive to the density and mass of the surface sites. Poblete et al.~found that increasing the surface-site density with a constant surface-site mass (matched to the mass of solvent in a collision cell \cite{ripoll_low-reynolds-number_2004, ripoll_dynamic_2005}) increased undesirable inertial effects \cite{poblete_hydrodynamics_2014}, but Peng and Sinno found that larger surface-site densities achieved better results for the long-time translational diffusion coefficient than smaller surface-site densities when the surface-site mass was held constant \cite{peng_multiparticle_2024}. Poblete et al.~also showed that particles with the same surface-site density but different surface-site masses had different linear velocity autocorrelation functions \cite{poblete_hydrodynamics_2014}. It is not currently well-established how to systematically and reliably select the surface-site density and mass.

Further, one strength of the discrete particle model in MPCD is that particles with complex shapes can be simulated \cite{wani_diffusion_2022,wani_diffusion_2022,rusenargun_influence_2023}, but modeling their potential energy function presents a challenge because their interactions are anisotropic (i.e., depend on both the relative position and orientation of the particles). Anisotropic interactions are mathematically difficult to formulate in terms of only position and orientation \cite{ramasubramani_mean-field_2020, fakhraei_approximation_2025, fakhraei_approximation_2026} and so are frequently approximated in simulations by discretizing the particle into many sites that have simpler pairwise isotropic interactions \cite{wani_diffusion_2022,wani_mesoscale_2024,rusenargun_influence_2023,kobayashi_structure_2020,kobayashi_structure_2020-1,kobayashi_self-assembly_2022}. Recently, one of us used such a model to simulate the diffusion and sedimentation of colloidal suspensions of hard cubes, octahedra, tetrahedra, and spherocylinders with MPCD \cite{wani_mesoscale_2024}. In that study, the same surface sites were used for both the particle interactions and for coupling the particles to the solvent, and it was noted that this choice produced a discrepancy between the hydrodynamic representation of the particle and the excluded volume of the particle because the pairwise site interactions (Weeks--Chandler--Andersen potential\cite{weeks_role_1971}) had a finite range of repulsion. This discrepancy had an observable impact on the structure of the tetrahedra at large particle volume fractions, but its impact on dynamic properties, particularly at lower concentrations and for other particle shapes, was less clear.

In this article, we investigate two aspects of discretizing colloidal particles in MPCD simulations. We first explore strategies for selecting the density and mass of surface sites used for coupling the particles to the solvent. We show that using a constant mass for the surface sites introduces a strong dependence of suspension transport properties on the surface-site density; however, choosing the site masses in a way that matches the mass and moment of inertia for a neutrally buoyant particle is both convenient and effective at removing the dependence of suspension transport properties on the surface-site density. We also show that this strategy generalizes across different solvents and different particles. We then investigate approaches for placing the sites that model the interactions between nearly-hard shape-anisotropic (polyhedral) colloidal particles. We propose using a second set of sites inside and tangent to the particle surface to effectively eliminate the discrepancy in the particle's shape between its hydrodynamic and excluded-volume interactions. We show that having such a discrepancy affects the suspension transport properties even at modest particle volume fractions.

The rest of the article is organized as follows. Section~\ref{sec:model} describes the models used for the solvent and colloidal particles, strategies for selecting the surface-site mass, and the methods employed for measuring suspension transport properties. Section~\ref{sec:results} critically interrogates hydrodynamic and excluded-volume discretization strategies for the colloidal particles. Section~\ref{sec:conclusions} summarizes our findings and recommendations.

\section{Model and Methods}
\label{sec:model}
Colloidal suspensions were simulated using an MPCD solvent with a discrete model for the colloidal particles \cite{poblete_hydrodynamics_2014,lobaskin_new_2004}. All quantities in this article are reported in a consistent system of units where $\ell$ is the unit of length, $m$ is the unit of mass, and $\varepsilon$ is the unit of energy. The unit of temperature is $\varepsilon/k_{\rm B}$, where $k_{\rm B}$ is the Boltzmann constant, and the unit of time is $\tau = \sqrt{m\ell^2/\varepsilon}$. All simulations were performed using HOOMD-blue \cite{anderson_hoomd-blue_2020,howard_efficient_2018,howard_efficient_2016,howard_quantized_2019} (version 6.1.1) extended with azplugins \cite{noauthor_mphowardlabazplugins_2026} (version 1.2.0).

\subsection{Solvent}
\label{sec:model:solvent}
Solvent particles with mass $1\,m$ were propagated in alternating streaming and collision steps. In the streaming step, the solvent particles moved according to Newton's equations of motion. In the collision step, the solvent particles were sorted into cubic cells with edge length $1\,\ell$ that were randomly shifted along each Cartesian axis by amounts drawn uniformly from $[-\ell/2,\ell/2]$ prior to every collision to ensure Galilean invariance \cite{ihle_stochastic_2003}. Solvent particles in the same cell then exchanged momentum according to the stochastic rotation dynamics (SRD) collision rule without angular momentum conservation using a rotation angle of $130^\circ$ about an axis randomly drawn from the unit sphere for each cell \cite{malevanets_mesoscopic_1999,allahyarov_mesoscopic_2002}. Previous studies have shown that neglect of angular momentum conservation in SRD is reasonable for bulk colloidal suspensions \cite{poblete_hydrodynamics_2014,yang_effect_2015}. A cell-level Maxwell--Boltzmann thermostat \cite{huang_cell-level_2010} was applied during the collision to maintain a constant temperature $T = 1\,\varepsilon/k_{\rm B}$.  The solvent mass density was $\rho_0 = 5\,m/\ell^3$, or 5 particles per cell on average, and collisions occurred every $0.1\,\tau$. These parameters give a solvent dynamic viscosity $\eta_0 = 3.96\,\varepsilon \tau/\ell^3$ \cite{statt_unexpected_2019,kikuchi_transport_2003,tuzel_transport_2003}.

\subsection{Colloidal particles}
The colloidal particles were coupled to the solvent using a discrete model \cite{poblete_hydrodynamics_2014}. The surface of a particle was discretized into sites with mass $m_{\rm s}$ that moved with the particle as a rigid body \cite{nguyen_rigid_2011} and participated in the collision step with the solvent using the procedure described in Ref.~\citenum{bush_simulating_2026}. Between collisions, the colloidal particles interacted with each other through pairwise potential energy functions and were propagated according to Newton's equations of motion using a quaternion-based velocity Verlet integration scheme \cite{miller_symplectic_2002}. We previously showed that this procedure can faithfully reproduce several transport properties of colloidal hard spheres \cite{bush_simulating_2026}.

We simulated colloidal suspensions of nearly hard spheres, cubes, regular octahedra, and regular tetrahedra. The integration timestep $\Delta t$ was $0.1\,\tau$ for the spheres and $0.05\,\tau$ for the regular polyhedra unless otherwise noted. We will discuss how the surface sites were generated and how the particles interactions were modeled for the spheres and for the regular polyhedra in Secs.~\ref{sec:model:colloids:spheres} and \ref{sec:model:colloids:poly}, respectively. Then, we will discuss the strategies we explored for choosing $m_{\rm s}$ in Sec.~\ref{sec:model:colloids:sitemass}.

\subsubsection{Spheres}
\label{sec:model:colloids:spheres}
We simulated spheres with diameter $d$ ranging from $3\,\ell$ to $12\,\ell$. The lower bound on $d$ was chosen because there are known artifacts in MPCD when the particle diameter becomes small compared to the size of the collision cell \cite{padding_hydrodynamic_2006}. The surface sites for the spheres were chosen from solutions to the Thomson problem \cite{wales_structure_2006}, which give good, uniform coverage of a sphere's surface and are publicly available \cite{wales_global_2006}. Using these solutions, rather than the vertices of a subdivided icosahedron as in previous work \cite{howard_transport_2026-1,kobayashi_structure_2020,kobayashi_structure_2020-1,kundu_exploring_2025,peng_multiparticle_2024,poblete_hydrodynamics_2014,wani_diffusion_2022,wani_mesoscale_2024,yetkin_structure_2024}, allowed us to systematically and incrementally vary the surface site density $\alpha$. We considered surface site densities that ranged from roughly $0.5\,\ell^{-2}$ to $3.0\,\ell^{-2}$ in steps of $0.5\,\ell^{-2}$ in order to test the influence of $\alpha$ on the transport properties of the suspension. The availability of reliable solutions to the Thomson problem for large numbers of points helped determine the upper bounds on $d$ and $\alpha$.

The spheres interacted with each other through the core-shifted Weeks--Chandler--Andersen potential \cite{weeks_role_1971},
\begin{equation}
u(r)= \begin{cases} 
    \displaystyle 4\varepsilon\left[ \left( \frac{\sigma}{r - \Delta} \right)^{12} -\left( \frac{\sigma}{r - \Delta}  \right)^{6} + \frac{1}{4} \right], & r\leq \Delta+2^{1/6}\sigma \\
    0, & \text{otherwise}
\end{cases},
\label{eq:WCA_potential}
\end{equation}
where $r$ is the center-to-center distance of the particles, $\sigma$ is the length scale of the repulsion, and $\Delta$ is the shift factor.  We used $\sigma = 1\,\ell$ and $\Delta = d-\sigma$, which one of us previously showed gives the expected isothermal compressibility for a fluid of hard spheres with $d$ ranging from $3\,\ell$ to $12\,\ell$ \cite{howard_transport_2026-1}. 

\subsubsection{Regular polyhedra}
\label{sec:model:colloids:poly}
We simulated three regular polyhedra---cubes, octahedra, and tetrahedra---with the same edge length, $a = 6\,\ell$. The surface sites were generated using a square mesh for the cubes and a mesh of equilateral triangles for the octahedra and tetrahedra with a chosen number of sites per edge $n_{\rm s}$. Using geometric arguments, the total number of surface sites per cube is $8 + 12(n_{\rm s} - 2) + 6(n_{\rm s} - 2)^2$, while the total number of surface sites per octahedron or tetrahedron is $n_{\rm v} + n_{\rm e} (n_{\rm s}-2) + n_{\rm f} (n_{\rm s}-2)(n_{\rm s}-3)/2$ where $n_{\rm v}$ is the number of vertices, $n_{\rm e}$ is the number of edges, and $n_{\rm f}$ is the number of faces for the particle. This discretization procedure is similar to that of Ref.~\citenum{wani_mesoscale_2024}, but it allows for arbitrary $n_{\rm s} \ge 2$. For the cubes, we varied $n_{\rm s}$ from 5 to 11 in order to vary $\alpha$ across a similar range as for the spheres; whereas, for the octahedra and tetrahedra, we fixed $n_{\rm s} = 9$ to achieve a surface density of roughly $2.0\,\ell^{-2}$.

\begin{figure}
  \centering
  \includegraphics{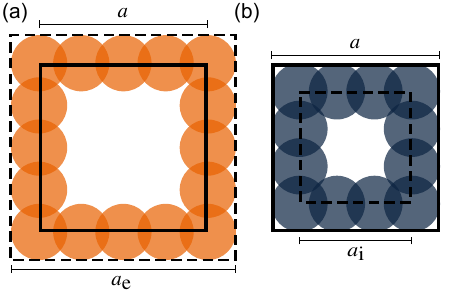}
  \caption{Schematic of placement of excluded-volume sites using (a) outer and (b) inner exclusion.}
  \label{fig:exclusion_schematic}
\end{figure}

The interactions between polyhedra were modeled using a discrete approach: the standard Weeks--Chandler--Anderson potential \cite{weeks_role_1971} [eq.~\eqref{eq:WCA_potential} with $\Delta = 0$] was applied between excluded-volume sites in different particles. In Refs. \citenum{wani_diffusion_2022} and \citenum{wani_mesoscale_2024}, Wani et al.~used the surface sites for coupling the polyhedra to the solvent as excluded-volume sites with $\sigma = 1.0\,\ell$. This choice was made for computational convenience, but it had the effect of increasing the excluded volume of the particles. Specifically, the edge length of the cube, octahedron, or tetrahedron that enclosed spheres of diameter $\sigma$ at the surface sites was $a_{\rm e} =a(1+\sigma/d_{\rm I})$, where $d_{\rm I}$ is the diameter of the inscribing sphere for the polyhedron with edge length $a$ [Fig.~\ref{fig:exclusion_schematic}(a)] \cite{wani_mesoscale_2024}. The excess excluded volume might be decreased by decreasing $\sigma$, but doing so requires a significant increase in the number of excluded-volume sites to still prevent particle overlap (Table S1). Hence, we also considered an alternative strategy that placed the excluded-volume sites on the surface of a polyhedron with smaller edge length $a_{\rm i} = a(1-\sigma/d_{\rm I})$, which ensures all spheres of diameter $\sigma$ are inside the polyhedron with edge length $a$ [Fig.~\ref{fig:exclusion_schematic}(b)]. We will refer to these two strategies for placing excluded-volume sites as outer exclusion and inner exclusion, respectively.

For both outer and inner exclusion, we considered two values of $\sigma$, $1.0\,\ell$ and $0.5\,\ell$. We set the number of excluded-volume sites per edge for outer exclusion so that the distance between them was $0.75\,\sigma$, giving 9 sites for $\sigma=1.0\,\ell$ and 17 sites for $\sigma=0.5\,\ell$. The number of sites per edge for inner exclusion case was one less than for outer exclusion. The total number of excluded-volume sites for each particle shape and discretization strategy is summarized in Table S1.

\subsubsection{Surface site mass}
\label{sec:model:colloids:sitemass}
Having described the colloidal particles, we return to the selection of the mass $m_{\rm s}$ of the surface sites used to couple them to the solvent. The conventional strategy for choosing $m_{\rm s}$ is to match the mass of a surface site to the average mass of solvent in a collision cell \cite{poblete_hydrodynamics_2014}, $m_{\rm s} = \rho_0 \ell^3$, which we will refer to as the C strategy. The C strategy can be traced to a foundational MPCD study by Ripoll et al. \cite{ripoll_dynamic_2005}, who found it balanced good hydrodynamic coupling against undesirable inertial effects for point-like solutes, and it has been used repeatedly in discrete models for colloidal particles \cite{poblete_hydrodynamics_2014,peng_multiparticle_2024,wani_diffusion_2022,wani_mesoscale_2024,kobayashi_structure_2020,kobayashi_structure_2020-1,kundu_exploring_2025,howard_transport_2026-1,das_clustering_2018,hu_modelling_2015,mauer_static_2017,myung_weak_2018,rusenargun_influence_2023,yetkin_structure_2024}. We note, however, that the C strategy necessarily leads to a dependence of the total mass $M$ and the diagonalized moment of inertia tensor $\mathbf{I}$ of the surface sites on their density $\alpha$, which may impact particle dynamics, because $m_{\rm s}$ is a constant that only depends on the solvent density. For example, larger $\alpha$ may lead to an undesirable increase in inertial effects, yet $\alpha$ must be sufficiently large to faithfully capture the shape of the surface and hydrodynamic interactions \cite{poblete_hydrodynamics_2014,peng_multiparticle_2024}. We hence sought an alternative strategy that achieved good hydrodynamic coupling for the colloid particles with less dependence on $\alpha$.

\begin{figure*}
  \centering
  \includegraphics{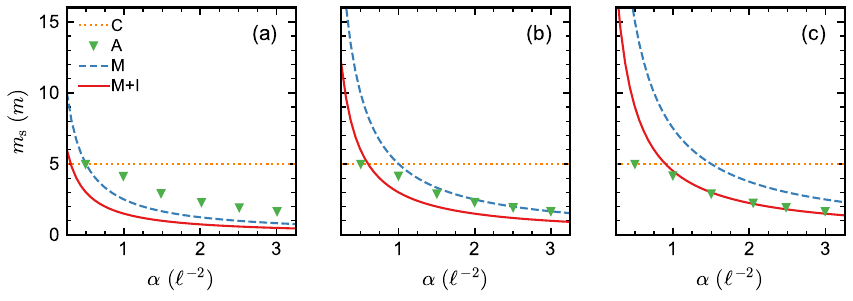}
  \caption{Mass of surface sites $m_{\rm s}$ as a function of surface-site density $\alpha$ for the C, A, M, and M+I strategies for spheres with diameter (a) $3\,\ell$, (b) $6\,\ell$, and (c) $9\,\ell$ using solvent density $\rho_0=5\,m/\ell^3$.}
  \label{fig:surface_mass}
\end{figure*}

A natural extension of the C strategy is to match the \textit{average} surface-site mass in a collision cell containing at least one surface site to the average mass of solvent in a collision cell, which we will refer to as the A strategy. In practice, the average number of surface sites in a collision cell must be calculated numerically. We carried out these calculations for the spheres using all pairwise combinations of 500 random orientations of the sphere and $10^3$ random shifts of the collision cells. The surface-site mass for the spheres using the A strategy appeared to be independent of colloid diameter (Fig.~\ref{fig:surface_mass}). However, we noted that $M$ and $\mathbf{I}$ still had a dependence on $\alpha$ using the A strategy, even though it was weaker than for the C strategy.

\begin{figure}
  \centering
  \includegraphics{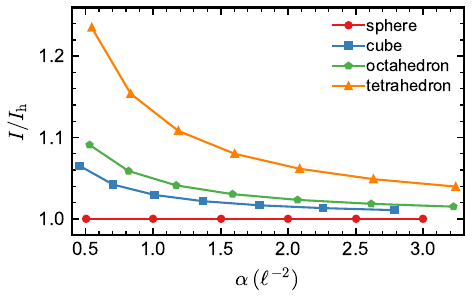}
  \caption{Moment of inertia $I$ for a sphere, cube, octahedron, and tetrahedron as a function of surface-site density $\alpha$ for the M strategy normalized by the moment of inertia of a hollow particle with the same total mass, $I_{\rm h}$.}
  \label{fig:inertia_shapes}
\end{figure}

The dependence of $M$ on $\alpha$ can be fully removed by choosing the surface-site mass so that the total mass of the sites matches a desired value, which we will refer to as the M strategy. A natural choice for total mass is that of a solid neutrally buoyant particle, $M_0 = \rho_0 V_0$ where $V_0$ is the volume of the particle. Distributing this mass equally between the surface sites gives $m_{\rm s} = M_0 / (\alpha A_0)$ where $A_0$ is the surface area of the particle; for example, $V_0 = \pi d^3/6$ and $A_0 = \pi d^2$ for a sphere, so $m_s=\rho_0 d/(6\alpha)$ (Fig.~\ref{fig:surface_mass}). For the M strategy, $m_{\rm s}$ must decrease as $\alpha$ increases in order to maintain constant $M$. A consequence of this choice is that the diagonalized moment of inertia tensor $\mathbf{I}$ approaches that of an equivalent hollow particle \cite{satterly_moments_1958} with the same mass $\mathbf{I}_{\rm h}$ as $\alpha$ increases. Indeed, for the sphere, we found that the averaged moment of inertia $I = {\rm tr}(\mathbf{I})/3$ was essentially that of a hollow sphere, $I_{\rm h} = M_0 d^2/6$, for all $\alpha$ (Fig.~\ref{fig:inertia_shapes}). For the regular polyhedra, the dependence of $I/I_{\rm h}$ on $\alpha$ was more complex, likely because of their anisotropy, but $I/I_{\rm h}$ also tended toward 1 as $\alpha$ increased.

It is possible to match both the mass $M_0$ and the moment of inertia $I_0$ of the solid neutrally buoyant particle, which we will refer to as the M+I strategy, by extending the M strategy. First, the surface-site mass $m_{\rm s}$ can be calculated so that $I = I_0$; for a solid sphere with $I_0 = M_0 d^2/10$, this procedure theoretically gives $m_s = I_0 / (\alpha A_0) = \rho_0 d/(10\alpha)$, which is less than for the M strategy regardless of $d$ (Fig.~\ref{fig:surface_mass}). Note that matching $\mathbf{I}$ to $\mathbf{I}_0$ using only $m_{\rm s}$ as a model parameter is not possible in general because these tensors may not be isotropic for some particles and discretizations, which is why we match the trace of $\mathbf{I}$. For a sphere, this procedure resulted in surface particles masses nearly identical to theoretical expectations. Second, the total mass of the surface sites is expected to be less than $M_0$ because it is effectively hollow, so additional mass can be inserted at the particle's center of mass to increase $M$ without changing $I$. This mass site was not coupled to the solvent and existed solely for matching $M_0$.

\subsection{Transport properties}
We measured the shear viscosity, sedimentation velocity, and long-time translational and rotational self-diffusion coefficients for suspensions with nominal volume fraction $\phi = N V_0/V$, where $N$ is the total number of particles and $V$ is the volume of the simulation box. The simulation protocols were similar to those described in Refs.~ \citenum{howard_transport_2026-1} and \citenum{bush_simulating_2026}. Equilibrated colloidal-particle configurations were generated using molecular dynamics simulations without any solvent, for which no surface sites were needed and the particle mass was hence chosen arbitrarily. For subsequent simulations with solvent, surface sites were added to the equilibrated particle configurations, and the mass and moment of inertia tensor for the particles was set based on $m_{\rm s}$. The uncertainties in these measurements were estimated as one standard error from 5 independent simulations.

Equilibrated configurations of spheres and cubes were generated in a cubic simulation box with edge length $L=120\,\ell$ and periodic boundary conditions to serve as initial configurations for simulations comparing strategies for selecting $\alpha$ and $m_{\rm s}$. Configurations were taken from Ref.~\citenum{howard_transport_2026-1} for the spheres with diameter $3\,\ell$, $6\,\ell$, and $12\,\ell$. For spheres with other diameters and cubes, isothermal--isochoric molecular dynamics simulations were conducted using an integration timestep $0.005\,\tau$ and a Bussi thermostat\cite{bussi_canonical_2007} with time constant $0.1\,\tau$ and temperature $1\,\varepsilon/k_{\rm B}$. Spheres with mass $5\,m$ were randomly placed into sites of a face-centered cubic lattice that was commensurate with the desired simulation box and did not have particle overlap. Cubes with mass $5\,m$ and the corresponding moment of inertia tensor for a solid were randomly placed into sites of a simple cubic lattice, and their interactions were modeled using inner exclusion with $\sigma=1.0\,\ell$. The particles were equilibrated for $10^3\,\tau$, then 5 configurations were sampled every $2 \times 10^4\,\tau$.

Additional molecular dynamics simulations were run for the polyhedra to not only generate initial configurations but also to compare different strategies for placing the excluded-volume sites. These simulations used an integration timestep of $0.05\,\tau$ and a Bussi thermostat with time constant of $1\,\tau$ and temperature $1\,\varepsilon/k_{\rm B}$. Polyhedra with masses and moment of inertia tensors for a neutrally buoyant solid with mass density $\rho_0$ were randomly placed into sites of a simple cubic lattice in a larger cubic simulation box with edge length $2L=240\,\ell$ that was then linearly reduced every $2.5\,\tau$ over a period of $500\,\tau$ until it reached $L = 120\,\ell$. The particles were then equilibrated for $10^3\,\tau$ before 5 configurations were sampled every $10^5\,\tau$ to serve as initial configurations for subsequent simulations. During this sampling, positions were also recorded every $10\,\tau$ for structural analysis, and the pressure $P$ was recorded every $1\,\tau$ to determine its average value. The particle positions were used to calculate the radial distribution function $g(r)$ for distances $r \le 25\,\ell$ using a histogram with bin width $0.1\,\ell$ and the static structure factor $S(q)$ for wavenumbers $2\pi/L \leq q \leq 40\pi/L$ with wavevector averaging in intervals of $2\pi/L$. These calculations were performed using freud (version 3.5.0) \cite{freud2020}. The static structure factor at zero wavenumber, $S(0) = \lim_{q\to 0} S(q)$ was then determined using a quadratic fit of $S(q)$ for $q \le 0.3\,\ell^{-1}$ for the cubes, $0.15\,\ell^{-1} \leq q \leq 0.40\,\ell^{-1}$ for the octahedra, and $0.15\,\ell^{-1} \leq q \leq 0.50\,\ell^{-1}$ for the tetrahedra.

The shear viscosity $\eta$ of the suspension was determined using reverse nonequilibrium simulations \cite{muller-plathe_reversing_1999,tenney_limitations_2010}. The equilibrated configuration was replicated to create an orthorhombic simulation box with length $240\,\ell$ in the $y$-direction, and the simulation box was filled with solvent. A bidirectional linear shear flow in the $x$-direction was generated by swapping velocities between pairs of solvent particles in regions of thickness $1.0\,\ell$ at the bottom and at the center of the simulation box. The particles in these regions were sorted based on how close the $x$-component of their velocities were to target values of $0.5\,\ell/\tau$ for the bottom region and  $-0.5\,\ell/\tau$ in the center region, then up to 100 pairs of sorted particles from each region exchanged the $x$-component of their velocities. After simulating for $2 \times 10^4\,\tau$ to allow the flow to develop, a simulation of $2 \times 10^4\,\tau$ was conducted to measure the cumulative momentum exchanged $p_x$ and the mass-averaged velocity of the suspension $u_x(y)$ in the $x$-direction as a function of the $y$ position. We recorded $p_x$ every $1\,\tau$ and confirmed it essentially increased at a constant rate, while $u_x(y)$ was calculated using a histogram with bin width $0.5\,\ell$ and samples taken every $10\,\tau$. The average rate of momentum transferred $\dot{p}_x$ was calculated from the difference in $p_x$ between the start and end of the simulation, and the shear rate $\dot{\gamma} = {\rm d}u_x/{\rm d}y$ was calculated from a linear regression of $u_x(y)$ excluding $12\,\ell$ from both sides of each region and using the symmetry of the shear flow. The shear viscosity was then calculated as $\eta = \dot{p}_x / (2 L^2 \dot{\gamma})$.

The sedimentation velocity of the colloidal particles $U$ was determined for a force of $F=0.5\,\varepsilon/\ell$ applied in the $x$-direction. The equilibrated particle configurations were solvated, and a counterforce was applied to all solvent particles to ensure the net force acting on the suspension was zero. The average velocity of all colloidal particles was recorded every $0.1\,\tau + \Delta t$. A correction
\begin{equation}
\Delta U = -\frac{F \Delta t}{2M} + \xi S(0) \frac{F}{6\pi\eta L},
\end{equation}
where $\xi = 2.837297$ and $\eta$ is the measured suspension shear viscosity, was added to account for measurement bias from recording the velocities at the end of an integration step \cite{bush_simulating_2026} (first term) and finite-size effects from the periodic boundary conditions \cite{mo_method_1994} (second term). For the spheres, we calculated $S(0)$ using the isothermal compressibility of the Carnahan--Starling equation of state for hard spheres \cite{carnahan_equation_1969, wani_diffusion_2022}, while for the polyhedra we used the simulated value of $S(0)$.

The long-time translational and rotational self-diffusion coefficients were determined by measuring the mean squared displacement (MSD) $\langle \Delta r^2 \rangle$ and mean squared angular displacement (MSAD) $\langle \Delta \varphi^2 \rangle$ in an equilibrium simulation. Solvent was added to the equilibrated configurations, and after an initial period of $10^3\,\tau$, the positions and orientations of the colloidal particles were sampled every $10\,\tau$ during a $10^5\,\tau$ simulation. A large number of simulations were performed to compare different strategies for choosing $m_{\rm s}$, so for these simulations, the coordinates of a maximum of $10^3$ colloidal particles were saved to limit the amount of data. The total rotational displacement vector $\Delta \boldsymbol{\varphi}$ was determined by accumulating the displacement $\Delta\boldsymbol{\varphi}_i$ of a unit vector $\hat{\mathbf{p}}_i$ rotating with the particle from configuration $i$ to the next configuration $i+1$ \cite{hunter_tracking_2011,kammerer_dynamics_1997},
\begin{equation}
\Delta \boldsymbol{\varphi}_i = \arccos(\hat{\mathbf{p}}_i \cdot \hat{\mathbf{p}}_{i+1}) (\hat{\mathbf{p}}_i \times \hat{\mathbf{p}}_{i+1}).
\end{equation}
We took the reference vector to be each of the Cartesian axes in the coordinate system where the moment of inertia tensor for the particles was diagonal, and we averaged over them. The MSD was calculated for times $t \le 2 \times 10^4\,\tau$ using all saved particles and the MSAD was calculated for $t \le 10^4$ using up to $10^3$ particles, taking all configurations as time origins for both calculations.
The long-time translational self-diffusion coefficient $D_{\rm T}$ was determined from the MSD,
\begin{equation}
D_{\rm T} = \lim_{t \to \infty} \frac{1}{6}\frac{{\rm d}\langle \Delta r^2\rangle}{{\rm d}t},
\end{equation}
while the long-time rotational self-diffusion coefficient $D_{\rm R}$ was determined using the MSAD \cite{kammerer_dynamics_1997},
\begin{equation}
D_{\rm R} = \lim_{t \to \infty} \frac{1}{4}\frac{{\rm d}\langle\Delta\varphi^2\rangle}{{\rm d}t}.
\end{equation}
The required derivatives were calculated numerically, and the limits were evaluated by averaging over $10^4\,\tau \le t \le 2 \times 10^4\,\tau$ for the MSD and $10^3\,\tau \le t \le 10^4\,\tau$ for the MSAD. A correction $\Delta D_{\rm T} = \xi k_{\rm B} T/(6\pi\eta L)$ was added to $D_{\rm T}$ to account for finite-size effects from the periodic boundary conditions\cite{dunweg_molecular_1993,yeh_system-size_2004}. Finite-size corrections have been developed for $D_{\rm R}$ in periodic boundary conditions \cite{vogele_finite-size-corrected_2019,linke_rotational_2018}, but the correction is negligible for the box size used.

\section{Results and Discussion}
\label{sec:results}

\subsection{Hydrodynamic discretization}
\begin{figure*}
  \centering
  \includegraphics{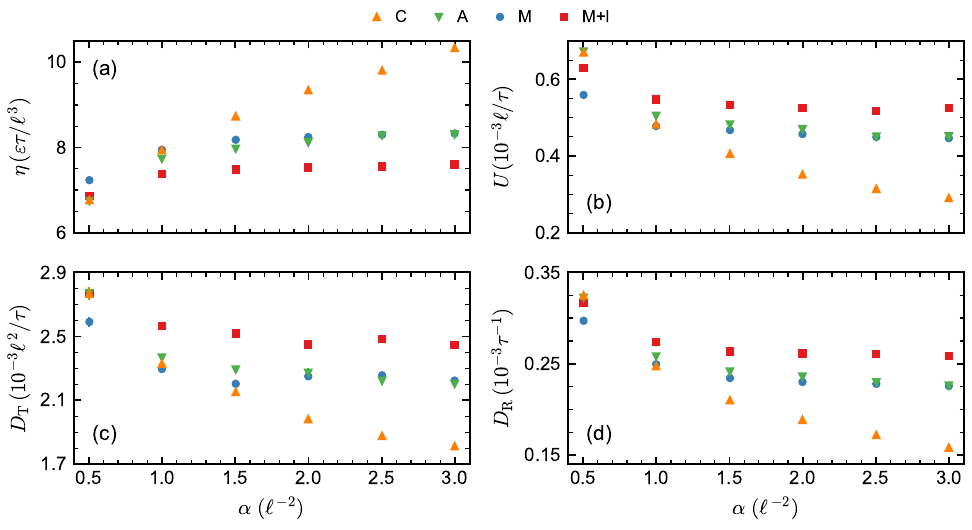}
  \caption{(a) Shear viscosity $\eta$, (b) sedimentation velocity $U$, (c) long-time translational self-diffusion coefficient $D_{\rm T}$, and (d) long-time rotational self-diffusion coefficient $D_{\rm R}$ for spheres with $d=6\,\ell$ when $\phi = 0.20$ using different surface-site densities $\alpha$ and strategies for selecting the surface-site mass $m_{\rm s}$.}
  \label{fig:6_sphere}
\end{figure*}

We first investigated strategies for selecting the density $\alpha$ and mass $m_{\rm s}$ of the surface sites that hydrodynamically couple the colloidal particles to the solvent. We started by considering the sphere with diameter $d = 6\,\ell$ at a volume fraction $\phi = 0.20$ (Fig.~\ref{fig:6_sphere}). This sphere was used in several previous studies \cite{howard_transport_2026-1,wani_diffusion_2022,wani_mesoscale_2024}, and its diameter is expected to be large enough relative to the collision cells to achieve good hydrodynamic coupling \cite{padding_hydrodynamic_2006}. For the C strategy, which is the current standard practice, all measured transport properties were strongly dependent on $\alpha$, with $\eta$ increasing and $U$, $D_{\rm T}$, and $D_{\rm R}$ decreasing as $\alpha$ increased from $0.5\,\ell^{-2}$ to $3.0\,\ell^{-2}$. In contrast, the transport properties for the A, M, and M+I strategies all reached a plateau within this range of $\alpha$. Similar results were obtained for spheres with diameter $3\,\ell$ (Fig.~S1) and $9\,\ell$ (Fig.~S2), as well as for spheres with diameter $6\,\ell$ in a solvent with a larger number density ($\rho_0 = 10\,m/\ell^3$, Fig.~S3) or using a different collision rule (Andersen thermostat \cite{allahyarov_mesoscopic_2002,noguchi_transport_2008}, Fig.~S4)

It is undesirable for the transport properties of the suspension to depend on $\alpha$ because it is a necessary but artificial model detail. The A, M, and M+I strategies were all favorable compared to the C strategy in this respect because they did not depend on the surface-site density for $\alpha \ge 2.0\,\ell^{-2}$. Comparing between these 3 strategies, the A and M strategies gave similar results across all transport properties for the spheres with $d=6\,\ell$, which makes sense because they had similar $m_{\rm s}$ [Fig.~\ref{fig:surface_mass}(b)]. The A and M+I strategies gave similar results for the spheres with $d=9\,\ell$ for the same reason. The M and M+I strategies were different from each other, despite giving the same total mass of the colloidal particle, because they had different moment of inertia tensors (giving different $m_{\rm s}$), and it is the surface sites that are responsible for the hydrodynamic coupling to the solvent. Practically, we found the A strategy was more cumbersome to use than the M and M+I strategies because it required numerical evaluation of the average number of surface sites in a cell. We hence chose to focus on the M and M+I strategies, which are related to straightforward properties of the colloidal particle, for further evaluation.

\begin{figure*}
  \centering
  \includegraphics{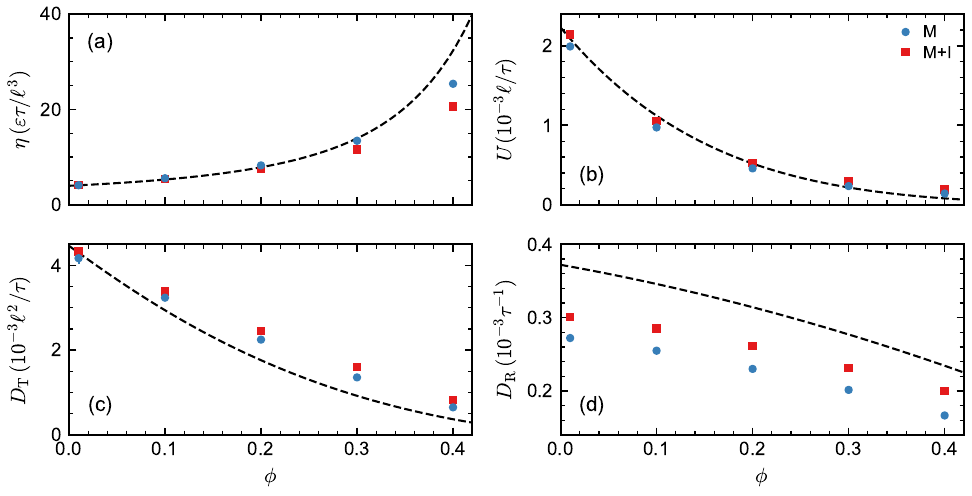}
  \caption{(a) Shear viscosity $\eta$, (b) sedimentation velocity $U$, (c) long-time translational self-diffusion coefficient $D_{\rm T}$, and (d) long-time rotational self-diffusion coefficient $D_{\rm R}$ for spheres with $d=6\,\ell$ at varying volume fraction $\phi$ using $\alpha=2.0\,\ell^{-2}$ with the M and M+I strategies for selecting the surface-site mass $m_{\rm s}$. The dashed black lines are theoretical predictions of (a) Eq.~\eqref{eq:viscosity}, (b) Eq.~\eqref{eq:U}, (c) Eq.~\eqref{eq:DT}, and (d) Eq.~\eqref{eq:DR}.}
  \label{fig:6_sphere_phi}
\end{figure*}

We fixed the surface-site density at $\alpha = 2.0\,\ell^{-2}$, and we computed transport properties for spheres with diameter $6\,\ell$ and volume fraction $\phi$ varying from 0.01 to 0.40 using the M and M+I strategies (Fig.~\ref{fig:6_sphere_phi}). Results for diameters $3\,\ell$ (Fig.~S5), $4\,\ell$ (Fig.~S6), and $9\,\ell$ (Fig.~S7) are available in the supplementary material. Analytical expressions have been derived for hard spheres for all transport coefficients that were measured as functions of $\phi$. The predicted shear viscosity, in the low-shear limit under which the simulations were conducted, is \cite{verberg_viscosity_1997}:
\begin{equation}
\frac{\eta}{\eta_0} = g^+\left[1 + \frac{1.44 (\phi g^+)^2}{1-0.1241\phi+10.46\phi^2} \right]
\label{eq:viscosity}
\end{equation}
where
\begin{equation}
g^+ = \frac{1-\phi/2}{(1-\phi)^3}
\end{equation}
is the value of the radial distribution function at contact modeled using the Carnahan--Starling equation of state \cite{carnahan_equation_1969}. The predicted sedimentation velocity is \cite{wang_short-time_2015-1}:
\begin{equation}
\frac{U}{U_0} = (1-\phi)^{6.5464},
\label{eq:U}
\end{equation}
where $U_0 = F/\gamma_{{\rm T},0}$ is the sedimentation velocity at infinite dilution and $\gamma_{{\rm T},0}$ is the single-particle translational friction coefficient ($3\pi\eta_0d$ for a sphere with no-slip boundary conditions). The predicted long-time translational self-diffusion coefficient is \cite{tokuyama_dynamics_1994}:
\begin{equation}
\frac{D_{\rm T}}{D_{{\rm T},0}} = \frac{1 - 9\phi/32}{1 + H + (\phi/\phi_0)/(1-\phi/\phi_0)^2},
\label{eq:DT}
\end{equation}
where $D_{{\rm T},0} = k_{\rm B}T/\gamma_{{\rm T},0}$ is the translational self-diffusion coefficient at infinite dilution and
\begin{equation}
H = \frac{2b^2}{1-b} - \frac{c}{1+2c} - \frac{b c (2+c)}{(1+c)(1-b+c)}
\end{equation}
with $b=\sqrt{9\phi/8}$, $c=11\phi/16$, and $\phi_0 \approx 0.5718$. Last, the predicted long-time rotational self-diffusion coefficient is \cite{clercx_three_1992}:
\begin{equation}
\frac{D_{\rm R}}{D_{{\rm R},0}} = 1 - 0.630\phi - 0.74\phi^2
\label{eq:DR}
\end{equation}
where $D_{{\rm R},0} = k_{\rm B} T/\gamma_{{\rm R},0}$ is the rotational self-diffusion coefficient at infinite dilution and $\gamma_{{\rm R},0}$ is the single-particle rotational friction coefficient ($\pi \eta_0 d^3$ for a sphere with no-slip boundary conditions). (An alternative formulation from Degiorgio et al. \cite{degiorgio_rotational_1995} uses $-0.67$ as the coefficient for $\phi^2$.) Equation \eqref{eq:DR} is valid for $\phi \leq 0.30$ and has been shown to overestimate $D_{\rm R}$ at higher concentrations \cite{hagen_rotational_1999,degiorgio_rotational_1995}.

Both the M and M+I strategies gave transport properties that were in qualitative agreement with theoretical expectations for $d = 6\,\ell$ (Fig.~\ref{fig:6_sphere_phi}). The closest quantitative agreement between simulation and theory was for $U$, while the biggest disagreement was for $D_{\rm R}$ with the simulations giving smaller values than theory. The M and M+I strategies gave similar results to each other for $\eta$, $U$, and $D_{\rm T}$, but gave different results for $D_{\rm R}$ with the M+I strategy being somewhat closer to theory for more dilute concentrations. Comparing between sphere diameters (Figs.~S5--S7), the differences between the M and M+I strategies, as well as the differences between the simulations and theory, seemed to be largest for $d=3\,\ell$ and to decrease as $d$ increased.

\begin{figure*}
  \centering
  \includegraphics{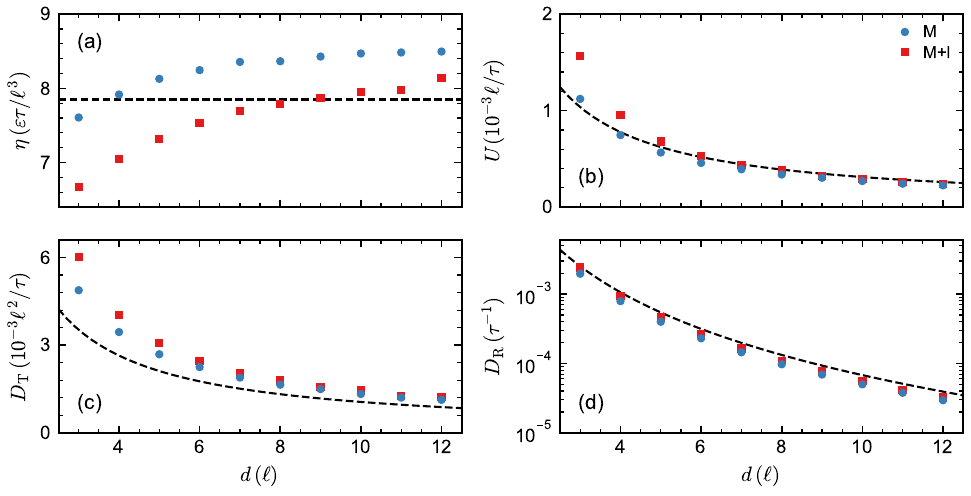}
  \caption{(a) Shear viscosity $\eta$, (b) sedimentation velocity $U$, (c) long-time translational self-diffusion coefficient $D_{\rm T}$, and (d) long-time rotational self-diffusion coefficient $D_{\rm R}$ for spheres with varying diameter $d$ when $\phi = 0.20$ using $\alpha=2.0\,\ell^{-2}$ with the M and M+I strategies for selecting the surface-site mass $m_{\rm s}$. The dashed black lines are theoretical predictions of (a) Eq.~\eqref{eq:viscosity}, (b) Eq.~\eqref{eq:U}, (c) Eq.~\eqref{eq:DT}, and (d) Eq.~\eqref{eq:DR}.}
  \label{fig:diameter}
\end{figure*}

To demonstrate more clearly, we fixed the volume fraction at $\phi = 0.20$ and varied $d$ from $3\,\ell$ to $12\,\ell$ with constant surface-site density $\alpha = 2.0\,\ell^{-2}$ (Fig.~\ref{fig:diameter}). All simulated properties appeared to come into better agreement with theoretical predictions as $d$ increased with the exception of $D_{\rm R}$, which had roughly the same agreement. The differences between the M and M+I strategies decreased for $U$ and $D_{\rm T}$ as $d$ increased but stayed roughly the same for $\eta$ and $D_{\rm R}$. These findings are consistent with prior work showing that particle diameters that are large compared to the collision cell are needed to achieve good hydrodynamic coupling. A previous study \cite{padding_hydrodynamic_2006} recommended a minimum diameter of $4\,\ell$; our results are in line with this finding but also reveal that a somewhat larger diameter may be needed.

Based on this analysis, we concluded that the M and M+I strategies both achieved the desired outcome: a reliable, simple strategy for selecting the surface-site density and mass. The M+I strategy was in somewhat better agreement with theory for $\eta$ and $D_{\rm R}$ for the spheres. It was also conceptually appealing that the M+I strategy allowed us to match both the total mass and moment of inertia of the solid particle. We confirmed that the M+I strategy gave consistent transport coefficients when $\alpha \ge 1.7\,\ell^2$ for cubes modeled using the inner exclusion strategy with $\sigma=1.0\,\ell$ at volume fraction $\phi = 0.20$ (Fig.~S8), supporting the potential generality of this strategy. We hence recommend using the M+I strategy to choose $m_{\rm S}$ and a surface-site density $\alpha \ge 2.0\,\ell^{-2}$.

To further test this recommendation, we simulated the diffusion and sedimentation of each particle studied at infinite dilution (Table \ref{tab:dilute_values}). We applied the same simulation protocols using a single particle in the simulation box, and we performed 100 simulations to measure the sedimentation velocity and 400 simulations to measure the long-time self-diffusion coefficients. The long-time limits of the derivative of the MSD and MSAD were obtained by averaging over the intervals $10^3\,\tau \le t \le 5 \times 10^3\,\tau$ and $5 \times 10^2\,\tau \le t \le 10^3\,\tau$, respectively. Uncertainties were estimated using block averaging with 5 blocks.

\begin{table*}
\caption{Sedimentation velocity $U_0$, long-time translational self-diffusion coefficient $D_{{\rm T},0}$, and long-time rotational self-diffusion coefficient $D_{{\rm R},0}$ of a single sphere with diameter $d = 6\,\ell$, cube, octahedron, or tetrahedron calculated from simulations (sim.) and theoretically estimated (est.).}
\label{tab:dilute_values}
\begin{tabular}{ccccccc}
 & \multicolumn{2}{c}{$U_0\,(10^{-3}\ell/\tau)$} & \multicolumn{2}{c}{$D_{{\rm T},0}\,(10^{-3} \ell^2 / \tau)$} & \multicolumn{2}{c}{$D_{{\rm R},0}\,(10^{-3}\,\tau^{-1})$} \\
 & sim. & est. & sim. & est. & sim. & est. \\
\hline
sphere & $2.22\pm0.06$ & 2.24 & $4.40\pm0.04$ & 4.48 & $0.31\pm0.00$ & 0.37 \\
cube & $1.61\pm0.02$ & 1.66 & $3.25\pm0.03$ & 3.33 & $0.12\pm0.00$ & 0.15 \\
octahedron & $2.18\pm0.02$ & 2.18 & $4.24\pm0.04$ & 4.36 & $0.27\pm0.00$ & 0.18 \\
tetrahedron & $3.40\pm0.07$ & 3.15 & $6.61\pm0.08$ & 6.29 & $0.80\pm0.00$ & 0.69
\end{tabular}
\end{table*}

Both $D_{{\rm T},0}$ and $U_0$ can be theoretically predicted from the single-particle translational friction coefficient, $\gamma_{{\rm T},0}$. This friction coefficient is well-known for a sphere, while Pettyjohn and Christiansen measured the settling rates of cubic, octahedral, and tetrahedral particles in a column at low Reynolds number \cite{pettyjohn_effect_1948} and found that they were correlated with the sphericity $\psi = \pi^{1/3} (6 V_0)^{2/3}/A_0$ \cite{wani_mesoscale_2024,pettyjohn_effect_1948},
\begin{equation}
\gamma_{{\rm T}, 0} = 3\pi \eta_0 \left( \frac{6V_0}{\pi} \right)^{1/3}\left[ 0.843 \log_{10}\left( \frac{\psi}{0.065} \right) \right]^{-1}.
\end{equation}
The single-particle rotational friction coefficient $\gamma_{{\rm R},0}$ is also well-known for spheres but must be estimated for the polyhedra. Okada and Satoh showed that $\gamma_{{\rm R}, 0}$ for a cube is roughly that of a sphere with a diameter equal to the mean of its inscribing and circumscribing spheres \cite{okada_evaluation_2020}. We will use this approach to estimate $\gamma_{{\rm R},0}$ for all the polyhedra.

We compared the simulated and estimated values of $U_0$, $D_{{\rm T},0}$, and $D_{{\rm R},0}$ (Table \ref{tab:dilute_values}), finding generally good agreement. The simulated values of $U_0$ and $D_{{\rm T},0}$ were in excellent agreement with relative error: $-0.9\%$ and $-1.8\%$ for the sphere, $-3.0\%$ and $-2.3\%$ for the cube, $0.0\%$ and $-2.7\%$ for the octahedron, and $8.0\%$ and $5.1\%$ for the tetrahedron. The simulated values of $D_{{\rm R},0}$ were in worse agreement, with the sphere and cube deviating in the negative direction by $-18.2\%$ and $-15.4\%$ while the octahedron and tetrahedron deviated in the positive direction by $52.8\%$ and $17.0\%$. This larger error in $D_{{\rm R},0}$ is not surprising given Fig.~\ref{fig:6_sphere_phi}, and we emphasize that the error is comparable to deviations reported in MPCD literature for other discretization schemes \cite{poblete_hydrodynamics_2014}. We also note that the agreement in $D_{{\rm T},0}$ for the octahedron and tetrahedron is a substantial improvement compared to Ref.~\citenum{wani_mesoscale_2024}. This result is significant because the hydrodynamic discretization used for the octahedron and tetrahedron here is the same as in Ref.~\citenum{wani_mesoscale_2024}, and the excluded-volume discretization is not relevant for a single particle. Hence, the only meaningful difference between the two simulation models is the mass and moment of inertia of the colloidal particle, showing that the change in surface-site mass (from the C strategy to the M+I strategy) led to better agreement with theory.

\subsection{Excluded-volume discretization}
\begin{figure*}
  \centering
  \includegraphics{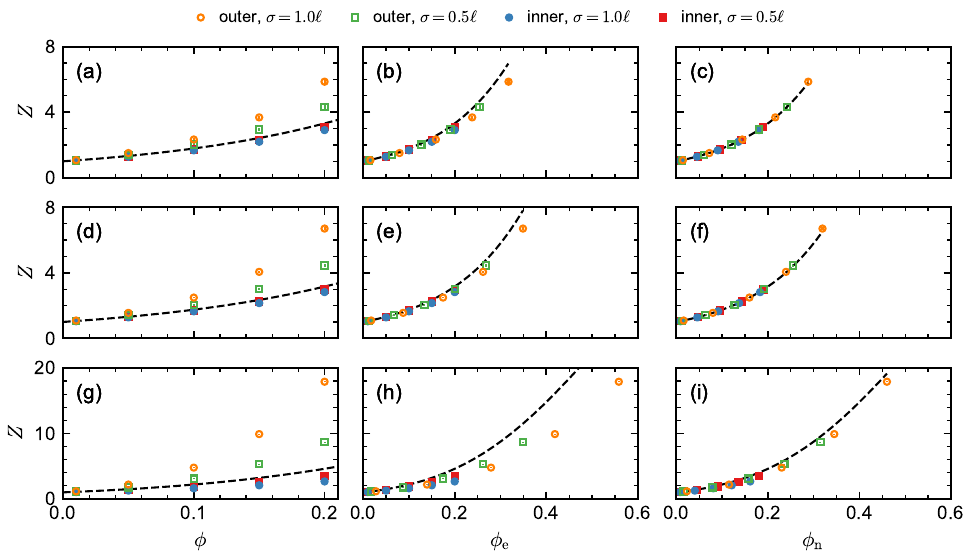}
  \caption{Compressibility factor $Z$ of (a--c) cubes, (d--f) octahedra, and (g--i) tetrahedra measured in isothermal--isochoric molecular dynamics simulations using outer and inner exclusion with $\sigma = 1.0\,\ell$ and $0.5\,\ell$ as a function of (a, d, g) nominal volume fraction $\phi$, (b, e, h) enclosing volume fraction $\phi_{\rm e}$, and (c, f, i) numerically estimated volume fraction $\phi_{\rm n}$. The dashed black lines are eq.~\eqref{eq:Z_eos} evaluated using the respective volume fractions.}
  \label{fig:Z}
\end{figure*}

Having established a good hydrodynamic discretization strategy, we next investigated strategies for placing sites to represent the excluded volume of the regular polyhedra. Differences in these strategies should manifest first in the thermodynamics and equilibrium structure of the suspension, so we started by measuring the compressibility factor $Z = P V/(N k_{\rm B} T)$ in isothermal--isochoric molecular dynamics simulations as a function of volume fraction $\phi$ (Fig.~\ref{fig:Z}). We compared the simulated values to theoretical expectations from an eighth-order virial expansion \cite{irrgang_virial_2017},
\begin{equation}
Z= 1 + \sum_{n=2}^8 B_n \phi^{n-1}
\label{eq:Z_eos}
\end{equation}
where $B_n$ is the $n$-th reduced virial coefficient. As anticipated, the simulated compressibility factor $Z$ depended strongly on the excluded-volume discretization. Outer exclusion gave values of $Z$ that were consistently larger than theory or inner exclusion, while inner exclusion consistently gave values of $Z$ that were less than theory. The deviations from theory were larger for outer exclusion than inner exclusion. Both exclusion strategies were also closer to theoretical expectations when $\sigma = 0.5\,\ell$ than when $\sigma = 1.0\,\ell$, which makes sense because the discretization is finer when $\sigma$ was smaller. The radial distribution functions (Figs.~S9--S11) showed a shift of the first peak to a longer distance and larger value for outer exclusion compared to inner exclusion, particularly for $\sigma = 1.0\,\ell$. Related behavior was found for the static structure factor $S(q)$, with a shift of its first peak to shorter wavelengths and larger values (Figs.~S12--S14) and a decrease in the extrapolated $S(0)$ (Fig.~S15) for outer exclusion compared to inner exclusion, again particularly for $\sigma = 1.0\,\ell$. All these results are consistent with outer exclusion generating a larger effective particle volume that depended on $\sigma$, while inner exclusion did not depend as strongly on $\sigma$.

To better understand the origin of these differences between strategies, we considered alternative definitions of the volume fraction based on different measures of the particle volume. In Ref. \citenum{wani_mesoscale_2024}, the particle volume was approximated as that of the polyhedron that enclosed spheres of diameter $\sigma$ at the excluded-volume sites, $V_{\rm e}$, giving an enclosing volume fraction $\phi_{\rm e} = (V_{\rm e}/V_0)\phi$ that is $\phi_{\rm e} = (1 + \sigma/d_{\rm I})^3 \phi$ for outer exclusion and $\phi_{\rm e} = \phi$ for inner exclusion. Using $\phi_{\rm e}$, we found good collapse of $Z$ for the cubes and octahedra [Figs.~\ref{fig:Z}(b) and \ref{fig:Z}(e)] but not the tetrahedra [Fig.~\ref{fig:Z}(h)], and all were systematically somewhat smaller than theoretical expectations. Analogous behavior was found comparing the extrapolated $S(0)$ to that predicted from the isothermal compressibility of the equation of state (Fig.~S15),
\begin{equation}
S(0)= \left( 1 + \sum_{n=2}^{8} n B_n \phi^{n-1} \right)^{-1}.
\label{eq:S_eos}
\end{equation}
Specifically, $S(0)$ was now systematically somewhat larger than theoretical expectations when plotted as a function of $\phi_{\rm e}$.

\begin{figure}
  \centering
  \includegraphics{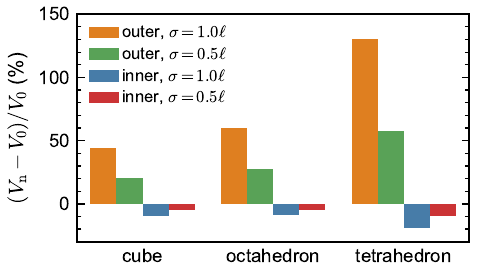}
  \caption{Percent difference between the numerically estimated volume $V_{\rm n}$ and the nominal volume $V_0$ of a cube, octahedron, and tetrahedron using outer or inner exclusion with $\sigma=1.0\,\ell$ or $0.5\,\ell$.}
  \label{fig:volume_diff}
\end{figure}

A possible reason for these discrepancies is that the enclosing polyhedron does not account for rounding or corrugation from using spheres to represent the particle. Accordingly, we numerically estimated the volume of the polyhedra using Monte Carlo integration. We randomly generated $10^7$ points in the volume between the polyhedron the excluded-volume sites laid on and the enclosing polyhedron using rejection sampling, then we determined the fraction of excluded volume in this space from the number of points that were inside the spheres of diameter $\sigma$ (Table S2). As expected, the numerically estimated volume $V_{\rm n}$ was larger than $V_0$ for outer exclusion and less than $V_0$ for inner exclusion, and the absolute percent difference was smaller when $\sigma$ was smaller (Fig.~\ref{fig:volume_diff}).  

Defining the numerically estimated volume fraction $\phi_{\rm n} = (V_{\rm n}/V_0) \phi$, we found that $Z$ for the cube and octahedron [Figs.~\ref{fig:Z}(c) and \ref{fig:Z}(f)] collapsed between discretization strategies and was in excellent agreement with theoretical expectations as a function of $\phi_{\rm n}$. The compressibility factor for the tetrahedron [Fig.~\ref{fig:Z}(i)] also reasonably collapsed and was in good, but not exact, agreement with theoretical expectations. Analogous improvements were obtained for $S(0)$ as a function of $\phi_{\rm n}$ (Fig.~S15). The differences for the tetrahedron are likely due to the difficulty in approximating the sharp edges and vertices of this shape using spheres. For example, the spheres create a rounding effect at the vertices similar to truncation, which is known to have an effect on packing behavior of tetrahedra \cite{damasceno_crystalline_2012} and virial coefficients \cite{irrgang_virial_2017}. For inner exclusion, the nearest distance from the surface of an excluded-volume sphere to a vertex of the polyhedron is $(\sigma/2)(d_{\rm C}/d_{\rm I} - 1)$, where $d_{\rm C}$ is the diameter of the circumscribing sphere for a polyhedron with edge length $a$. This formula simplifies to $\sigma(\sqrt{3} - 1)/2 \approx 0.366\sigma$ for the cube and octahedron but $\sigma$ for the tetrahedron, emphasizing the larger significance of rounding for the tetrahedron. This effect manifests in the larger relative difference in $V_{\rm n}$ for the tetrahedron with inner exclusion than for the cube or octahedron.

Practically, we desire an excluded-volume discretization that gives the most faithful representation of the nominal particle shape because the hydrodynamic discretization uses the nominal particle shape to place the surface sites. We hence recommend inner exclusion over outer exclusion, with $\sigma$ chosen to be as large as possible to reduce the number of pairwise interactions evaluated between excluded-volume sites. Based on Figs.~\ref{fig:Z}, \ref{fig:volume_diff}, and S15, we chose to use inner exclusion with $\sigma = 1.0\,\ell$ for the cubes and octahedra and $\sigma = 0.5\,\ell$ for the tetrahedra. These values of $\sigma$ were chosen because they lead to comparable accuracy in $V_{\rm n}$ (about 10\% error), $Z$, and $S(0)$. We used these models to simulate suspension transport properties for the polyhedra for volume fractions $0.01 \le \phi \le 0.20$.

\begin{figure*}
  \centering
  \includegraphics{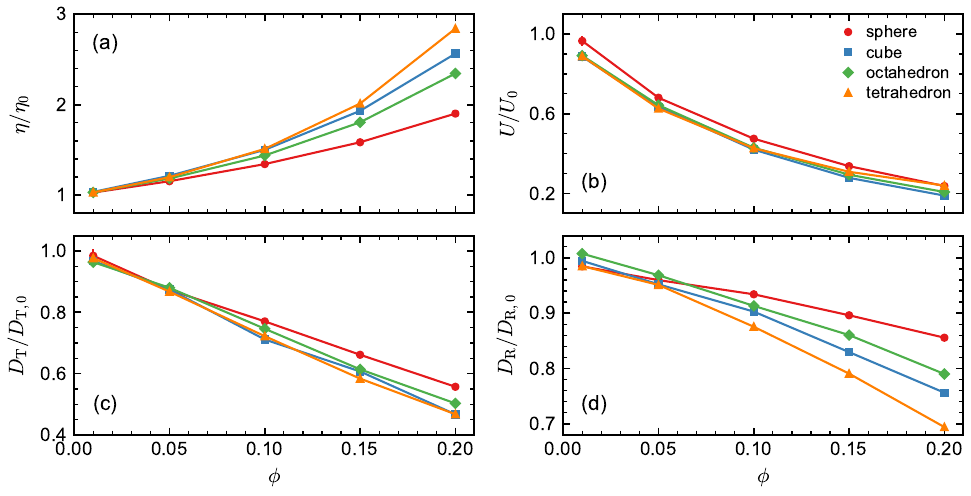}
  \caption{(a) Shear viscosity $\eta$, (b) sedimentation velocity $U$, (c) long-time translational self-diffusion coefficient $D_{\rm T}$, and (d) long-time rotational self-diffusion coefficient $D_{\rm R}$ normalized by their respective values at infinite dilution for spheres with $d = 6\,\ell$, cubes, octahedra, and tetrahedra as a function of volume fraction $\phi$. The excluded-volume for the polyhedra was represented using inner exclusion with $\sigma = 1.0\,\ell$ for the cubes and octahedra and $\sigma = 0.5\,\ell$ for the tetrahedra.}
  \label{fig:normed_phi}
\end{figure*}

To compare results between the different particle shapes, we normalized the suspension transport properties by their values at infinite dilution, and we included results for a sphere with diameter $6\,\ell$ with the polyhedra (Fig.~\ref{fig:normed_phi}). The directly measured values of the transport properties are available in Fig.~S16. We found that the sedimentation velocity $U$ and the long-time translational self-diffusion coefficient $D_{\rm T}$ had a similar dependence on $\phi$ regardless of shape, although the sphere had a somewhat weaker dependence for $D_{\rm T}$. On the other hand, we found that the shear viscosity $\eta$ and the long-time rotational self-diffusion coefficient $D_{\rm R}$ had a different dependence on $\phi$ for the different particle shapes. The spheres had the weakest dependence of $\eta$ and $D_{\rm R}$ on $\phi$, while the tetrahedra had the strongest dependence.

The dependence of $U$ and $D_{\rm T}$ on $\phi$ for the polyhedra was qualitatively different from what was found in Ref.~\citenum{wani_mesoscale_2024}, where the $\phi$-dependence of these quantities tended to be stronger and was more dissimilar between shapes. This difference is likely the result of the excluded-volume discretization: Ref.~\citenum{wani_mesoscale_2024} used a strategy similar to outer exclusion with $\sigma = 1.0\,\ell$. To test this hypothesis, we conducted additional simulations of the polyhedra using outer exclusion with $\sigma=1.0\,\ell$ (Fig.~S17). We did not include the tetrahedra at $\phi = 0.20$ in these simulations because $\phi_{\rm n} = 0.47$ for outer exclusion with $\sigma=1.0\,\ell$, and the tetrahedra were expected to undergo a phase transition \cite{haji-akbari_disordered_2009,haji-akbari_phase_2011}. Consistent with our hypothesis and Ref.~\citenum{wani_mesoscale_2024}, we found that the transport properties for the polyhedra no longer collapsed as a function of $\phi$, and the tetrahedra had the strongest $\phi$-dependence. Unlike the compressibility factor (Fig.~\ref{fig:Z}), the transport properties could not be collapsed using $\phi_{\rm n}$ [Fig.~S17(b, e, h, k)], but reasonable collapse was found using the average of $\phi$ and $\phi_{\rm n}$ [Fig.~S17(c, f, i, l)]. These observations emphasize that differences in the particle shape associated with the hydrodynamic discretization and the excluded-volume discretization can lead to significant differences in transport properties, and hence, an excluded-volume discretization that leads to a large discrepancy from the nominal particle volume $V_0$ is not recommended.

\section{Conclusions}
\label{sec:conclusions}
We have explored different strategies for discretizing colloidal particles for MPCD simulations. We first considered different approaches for choosing the density and the mass of surface sites used to hydrodynamically couple the colloidal particles to the solvent. We found that the conventional strategy of setting the site mass equal to the average mass of solvent in a collision cell gave suspension transport properties that were highly sensitive to the surface density, but we showed that consistent suspension transport properties could be obtained at sufficiently large site densities by scaling the site mass. Of the strategies considered, we recommend using a minimum surface site density of $2.0\,\ell^{-2}$ and matching the total mass and moment of inertia for a neutrally buoyant solid particle because this strategy is directly tied to physical properties of the colloidal particle and simple to implement. We then explored the implications of using different approaches to represent the excluded volume of nearly-hard shape-anisotropic colloidal particles using a collection of interaction sites. We found that having a discrepancy between the actual excluded volume and the nominal excluded volume led to a different volume-fraction dependence of the suspension transport properties. We hence recommend employing a second set of interaction sites, distinct from the surface sites used for the hydrodynamic discretization, that lie inside and tangent to the particle surface to represent the excluded volume. Both these recommendations are more easily implemented using rigid bodies than harmonic bonds to maintain the site geometry, and we recently developed a method for coupling rigid bodies to the MPCD solvent that is available in HOOMD-blue \cite{bush_simulating_2026}.

\section*{Supplementary Material}
See the supplementary material for transport properties of spheres with different sizes as a function of surface-site density and volume fraction, of spheres in solvents with different parameters as a function of surface-site density, and of cubes as a function of surface-site density; radial distribution functions, static structure factors, and transport properties for regular polyhedra as a function of volume fraction; comparison of transport properties for regular polyhedra with inner and outer exclusion as a function of volume fraction; and additional information about the particle discretization and volume.

\section*{Conflicts of interest}
The authors have no conflicts to disclose.

\section*{Data Availability}
The data that support the findings of this study are available from the authors upon reasonable request.

\section*{Acknowledgments}
This material is based upon work supported by the National Science Foundation under Award No.~2310724. This work used Delta at the National Center for Supercomputing Applications through allocation CHM250060 from the Advanced Cyberinfrastructure Coordination Ecosystem: Services \& Support (ACCESS) program \cite{access}, which is supported by National Science Foundation grants \#2138259, \#2138286, \#2138307, \#2137603, and \#2138296.

\bibliography{references}

@inproceedings{access,
  title = {{{ACCESS}}: {{Advancing}} Innovation: {{NSF}}'s Advanced Cyberinfrastructure Coordination Ecosystem: {{Services}} \& Support},
  booktitle = {Practice and Experience in Advanced Research Computing 2023: {{Computing}} for the Common Good},
  author = {Boerner, Timothy J. and Deems, Stephen and Furlani, Thomas R. and Knuth, Shelley L. and Towns, John},
  year = 2023,
  series = {{{PEARC}} '23},
  pages = {173--176},
  publisher = {Association for Computing Machinery},
  address = {New York, NY, USA},
  doi = {10.1145/3569951.3597559},
  isbn = {978-1-4503-9985-2}
}

@article{allahyarov_mesoscopic_2002,
  title = {Mesoscopic Solvent Simulations: {{Multiparticle-collision}} Dynamics of Three-Dimensional Flows},
  shorttitle = {Mesoscopic Solvent Simulations},
  author = {Allahyarov, E. and Gompper, G.},
  year = 2002,
  month = sep,
  journal = {Physical Review E},
  volume = {66},
  number = {3},
  pages = {036702},
  publisher = {American Physical Society},
  doi = {10.1103/PhysRevE.66.036702},
  urldate = {2025-11-17}
}

@article{anderson_hoomd-blue_2020,
  title = {{{HOOMD-blue}}: {{A Python}} Package for High-Performance Molecular Dynamics and Hard Particle {{Monte Carlo}} Simulations},
  shorttitle = {{{HOOMD-blue}}},
  author = {Anderson, Joshua A. and Glaser, Jens and Glotzer, Sharon C.},
  year = 2020,
  month = feb,
  journal = {Computational Materials Science},
  volume = {173},
  pages = {109363},
  issn = {0927-0256},
  doi = {10.1016/j.commatsci.2019.109363},
  urldate = {2025-10-14}
}

@article{boles_self-assembly_2016,
  title = {Self-{{Assembly}} of {{Colloidal Nanocrystals}}: {{From Intricate Structures}} to {{Functional Materials}}},
  shorttitle = {Self-{{Assembly}} of {{Colloidal Nanocrystals}}},
  author = {Boles, Michael A. and Engel, Michael and Talapin, Dmitri V.},
  year = 2016,
  month = sep,
  journal = {Chemical Reviews},
  volume = {116},
  number = {18},
  pages = {11220--11289},
  publisher = {American Chemical Society},
  issn = {0009-2665},
  doi = {10.1021/acs.chemrev.6b00196},
  urldate = {2026-04-09}
}

@article{brady_stokesian_1988,
  title = {Stokesian {{Dynamics}}},
  author = {Brady, J. F. and Bossis, G.},
  year = 1988,
  month = jan,
  journal = {Annual Review of Fluid Mechanics},
  volume = {20},
  number = {Volume 20, 1988},
  pages = {111--157},
  publisher = {Annual Reviews},
  issn = {0066-4189, 1545-4479},
  doi = {10.1146/annurev.fl.20.010188.000551},
  urldate = {2026-03-12},
  langid = {english}
}

@article{bush_simulating_2026,
  title = {Simulating Hydrodynamic Interactions in Colloidal Suspensions Using Multiparticle Collision Dynamics with Rigid-Body Constraints},
  author = {Bush, Michaela and Palmer, Jeremy C. and Howard, Michael P.},
  year = 2026,
  month = jul,
  journal = {The Journal of Chemical Physics},
  volume = {165},
  number = {4},
  pages = {044904},
  issn = {0021-9606},
  doi = {10.1063/5.0339394},
  urldate = {2026-07-23}
}

@article{bussi_canonical_2007,
  title = {Canonical Sampling through Velocity Rescaling},
  author = {Bussi, Giovanni and Donadio, Davide and Parrinello, Michele},
  year = 2007,
  month = jan,
  journal = {The Journal of Chemical Physics},
  volume = {126},
  number = {1},
  pages = {014101},
  issn = {0021-9606},
  doi = {10.1063/1.2408420},
  urldate = {2026-06-26}
}

@article{carnahan_equation_1969,
  title = {Equation of {{State}} for {{Nonattracting Rigid Spheres}}},
  author = {Carnahan, Norman F. and Starling, Kenneth E.},
  year = 1969,
  month = jul,
  journal = {The Journal of Chemical Physics},
  volume = {51},
  number = {2},
  pages = {635--636},
  issn = {0021-9606},
  doi = {10.1063/1.1672048},
  urldate = {2026-04-08}
}

@article{chen_lattice_1998,
  title = {{{LATTICE BOLTZMANN METHOD FOR FLUID FLOWS}}},
  author = {Chen, Shiyi and Doolen, Gary D.},
  year = 1998,
  month = jan,
  journal = {Annual Review of Fluid Mechanics},
  volume = {30},
  number = {Volume 30, 1998},
  pages = {329--364},
  publisher = {Annual Reviews},
  issn = {0066-4189, 1545-4479},
  doi = {10.1146/annurev.fluid.30.1.329},
  urldate = {2025-11-18},
  langid = {english}
}

@article{clercx_three_1992,
  title = {Three Particle Hydrodynamic Interactions in Suspensions},
  author = {Clercx, H. J. H. and Schram, P. P. J. M.},
  year = 1992,
  month = feb,
  journal = {The Journal of Chemical Physics},
  volume = {96},
  number = {4},
  pages = {3137--3151},
  issn = {0021-9606},
  doi = {10.1063/1.462843},
  urldate = {2026-05-18}
}

@article{damasceno_crystalline_2012,
  title = {Crystalline {{Assemblies}} and {{Densest Packings}} of a {{Family}} of {{Truncated Tetrahedra}} and the {{Role}} of {{Directional Entropic Forces}}},
  author = {Damasceno, Pablo F. and Engel, Michael and Glotzer, Sharon C.},
  year = 2012,
  month = jan,
  journal = {ACS Nano},
  volume = {6},
  number = {1},
  pages = {609--614},
  publisher = {American Chemical Society},
  issn = {1936-0851},
  doi = {10.1021/nn204012y},
  urldate = {2026-06-30}
}

@article{das_clustering_2018,
  title = {Clustering and Dynamics of Particles in Dispersions with Competing Interactions: Theory and Simulation},
  shorttitle = {Clustering and Dynamics of Particles in Dispersions with Competing Interactions},
  author = {Das, Shibananda and Riest, Jonas and Winkler, Roland G. and Gompper, Gerhard and Dhont, Jan K. G. and N{\"a}gele, Gerhard},
  year = 2018,
  journal = {Soft Matter},
  volume = {14},
  number = {1},
  pages = {92--103},
  issn = {1744-683X, 1744-6848},
  doi = {10.1039/C7SM02019H},
  urldate = {2025-09-16},
  langid = {english}
}

@article{degiorgio_rotational_1995,
  title = {Rotational Diffusion in Concentrated Colloidal Dispersions of Hard Spheres},
  author = {Degiorgio, Vittorio and Piazza, Roberto and Jones, Robert B.},
  year = 1995,
  month = sep,
  journal = {Physical Review E},
  volume = {52},
  number = {3},
  pages = {2707--2717},
  publisher = {American Physical Society},
  doi = {10.1103/PhysRevE.52.2707},
  urldate = {2026-05-18}
}

@article{dunweg_molecular_1993,
  title = {Molecular Dynamics Simulation of a Polymer Chain in Solution},
  author = {D{\"u}nweg, Burkhard and Kremer, Kurt},
  year = 1993,
  month = nov,
  journal = {The Journal of Chemical Physics},
  volume = {99},
  number = {9},
  pages = {6983--6997},
  issn = {0021-9606},
  doi = {10.1063/1.465445},
  urldate = {2025-10-06}
}

@article{fakhraei_approximation_2025,
  title = {Approximation of Anisotropic Pair Potentials Using Multivariate Interpolation},
  author = {Fakhraei, Mohammadreza and Kieslich, Chris A. and Howard, Michael P.},
  year = 2025,
  month = jul,
  journal = {The Journal of Physical Chemistry. B},
  volume = {129},
  number = {27},
  pages = {6985--6996},
  issn = {1520-6106},
  doi = {10.1021/acs.jpcb.5c01451},
  urldate = {2026-08-31},
  pmcid = {PMC12444758},
  pmid = {40576407}
}

@article{fakhraei_approximation_2026,
  title = {Approximation of Forces and Torques from Anisotropic Pairwise Interactions Using Multivariate Polynomials},
  author = {Fakhraei, Mohammadreza and Bush, Michaela and Kieslich, Chris A. and Howard, Michael P.},
  year = 2026,
  month = apr,
  journal = {The Journal of Chemical Physics},
  volume = {164},
  number = {14},
  pages = {144117},
  issn = {0021-9606},
  doi = {10.1063/5.0318270},
  urldate = {2026-05-04}
}

@article{freud2020,
  title = {Freud: A Software Suite for High Throughput Analysis of Particle Simulation Data},
  author = {Ramasubramani, Vyas and Dice, Bradley D. and Harper, Eric S. and Spellings, Matthew P. and Anderson, Joshua A. and Glotzer, Sharon C.},
  year = 2020,
  journal = {Computer Physics Communications},
  volume = {254},
  pages = {107275},
  issn = {0010-4655},
  doi = {10.1016/j.cpc.2020.107275}
}

@incollection{gompper_multi-particle_2009,
  title = {Multi-{{Particle Collision Dynamics}}: {{A Particle-Based Mesoscale Simulation Approach}} to the {{Hydrodynamics}} of {{Complex Fluids}}},
  shorttitle = {Multi-{{Particle Collision Dynamics}}},
  booktitle = {Advanced {{Computer Simulation Approaches}} for {{Soft Matter Sciences III}}},
  author = {Gompper, G. and Ihle, T. and Kroll, D. M. and Winkler, R. G.},
  editor = {Holm, Christian and Kremer, Kurt},
  year = 2009,
  pages = {1--87},
  publisher = {Springer},
  address = {Berlin, Heidelberg},
  doi = {10.1007/978-3-540-87706-6_1},
  urldate = {2024-12-11},
  isbn = {978-3-540-87706-6},
  langid = {english}
}

@article{hagen_rotational_1999,
  title = {Rotational Diffusion in Dense Suspensions},
  author = {Hagen, M. H. J and Frenkel, D and Lowe, C. P},
  year = 1999,
  month = oct,
  journal = {Physica A: Statistical Mechanics and its Applications},
  volume = {272},
  number = {3},
  pages = {376--391},
  issn = {0378-4371},
  doi = {10.1016/S0378-4371(99)00283-6},
  urldate = {2026-04-15}
}

@article{haji-akbari_disordered_2009,
  title = {Disordered, Quasicrystalline and Crystalline Phases of Densely Packed Tetrahedra},
  author = {{Haji-Akbari}, Amir and Engel, Michael and Keys, Aaron S. and Zheng, Xiaoyu and Petschek, Rolfe G. and {Palffy-Muhoray}, Peter and Glotzer, Sharon C.},
  year = 2009,
  month = dec,
  journal = {Nature},
  volume = {462},
  number = {7274},
  pages = {773--777},
  issn = {1476-4687},
  doi = {10.1038/nature08641},
  langid = {english},
  pmid = {20010683}
}

@article{haji-akbari_phase_2011,
  title = {Phase Diagram of Hard Tetrahedra},
  author = {{Haji-Akbari}, Amir and Engel, Michael and Glotzer, Sharon C.},
  year = 2011,
  month = nov,
  journal = {The Journal of Chemical Physics},
  volume = {135},
  number = {19},
  pages = {194101},
  issn = {0021-9606},
  doi = {10.1063/1.3651370},
  urldate = {2026-05-13}
}

@article{howard_efficient_2016,
  title = {Efficient Neighbor List Calculation for Molecular Simulation of Colloidal Systems Using Graphics Processing Units},
  author = {Howard, Michael P. and Anderson, Joshua A. and Nikoubashman, Arash and Glotzer, Sharon C. and Panagiotopoulos, Athanassios Z.},
  year = 2016,
  month = jun,
  journal = {Computer Physics Communications},
  volume = {203},
  pages = {45--52},
  issn = {0010-4655},
  doi = {10.1016/j.cpc.2016.02.003},
  urldate = {2026-08-31}
}

@article{howard_efficient_2018,
  title = {Efficient Mesoscale Hydrodynamics: {{Multiparticle}} Collision Dynamics with Massively Parallel {{GPU}} Acceleration},
  shorttitle = {Efficient Mesoscale Hydrodynamics},
  author = {Howard, Michael P. and Panagiotopoulos, Athanassios Z. and Nikoubashman, Arash},
  year = 2018,
  month = sep,
  journal = {Computer Physics Communications},
  volume = {230},
  pages = {10--20},
  issn = {0010-4655},
  doi = {10.1016/j.cpc.2018.04.009},
  urldate = {2025-04-23}
}

@article{howard_evaporation-induced_2018,
  title = {Evaporation-Induced Assembly of Colloidal Crystals},
  author = {Howard, Michael P. and Reinhart, Wesley F. and Sanyal, Tanmoy and Shell, M. Scott and Nikoubashman, Arash and Panagiotopoulos, Athanassios Z.},
  year = 2018,
  month = sep,
  journal = {The Journal of Chemical Physics},
  volume = {149},
  number = {9},
  pages = {094901},
  issn = {0021-9606},
  doi = {10.1063/1.5043401},
  urldate = {2026-06-30}
}

@article{howard_modeling_2019,
  title = {Modeling Hydrodynamic Interactions in Soft Materials with Multiparticle Collision Dynamics},
  author = {Howard, Michael P and Nikoubashman, Arash and Palmer, Jeremy C},
  year = 2019,
  month = mar,
  journal = {Current Opinion in Chemical Engineering},
  series = {Frontiers of {{Chemical Engineering}}: {{Molecular Modeling}}},
  volume = {23},
  pages = {34--43},
  issn = {2211-3398},
  doi = {10.1016/j.coche.2019.02.007},
  urldate = {2024-12-11}
}

@article{howard_quantized_2019,
  title = {Quantized Bounding Volume Hierarchies for Neighbor Search in Molecular Simulations on Graphics Processing Units},
  author = {Howard, Michael P. and Statt, Antonia and Madutsa, Felix and Truskett, Thomas M. and Panagiotopoulos, Athanassios Z.},
  year = 2019,
  month = jun,
  journal = {Computational Materials Science},
  volume = {164},
  pages = {139--146},
  issn = {0927-0256},
  doi = {10.1016/j.commatsci.2019.04.004},
  urldate = {2026-04-14}
}

@article{howard_transport_2026-1,
  title = {Transport Properties of Monodisperse and Bidisperse Hard-Sphere Colloidal Suspensions from Multiparticle Collision Dynamics Simulations},
  author = {Howard, Michael P.},
  year = 2026,
  month = may,
  journal = {The Journal of Chemical Physics},
  volume = {164},
  number = {19},
  pages = {194901},
  issn = {0021-9606},
  doi = {10.1063/5.0332320},
  urldate = {2026-06-26}
}

@article{hu_modelling_2015,
  title = {Modelling the Mechanics and Hydrodynamics of Swimming {{E}}. Coli},
  author = {Hu, Jinglei and Yang, Mingcheng and Gompper, Gerhard and G.~Winkler, Roland},
  year = 2015,
  journal = {Soft Matter},
  volume = {11},
  number = {40},
  pages = {7867--7876},
  publisher = {Royal Society of Chemistry},
  doi = {10.1039/C5SM01678A},
  urldate = {2025-09-04},
  langid = {english}
}

@article{huang_cell-level_2010,
  title = {Cell-Level Canonical Sampling by Velocity Scaling for Multiparticle Collision Dynamics Simulations},
  author = {Huang, C. C. and Chatterji, A. and Sutmann, G. and Gompper, G. and Winkler, R. G.},
  year = 2010,
  month = jan,
  journal = {Journal of Computational Physics},
  volume = {229},
  number = {1},
  pages = {168--177},
  issn = {0021-9991},
  doi = {10.1016/j.jcp.2009.09.024},
  urldate = {2025-08-08}
}

@article{HUMP96,
  title = {{{VMD}} -- {{Visual Molecular Dynamics}}},
  author = {Humphrey, William and Dalke, Andrew and Schulten, Klaus},
  year = 1996,
  journal = {Journal of Molecular Graphics},
  volume = {14},
  pages = {33--38},
  tbreference = {222},
  tbstatus = {Published.}
}

@article{hunter_tracking_2011,
  title = {Tracking Rotational Diffusion of Colloidal Clusters},
  author = {Hunter, Gary L. and Edmond, Kazem V. and Elsesser, Mark T. and Weeks, Eric R.},
  year = 2011,
  month = aug,
  journal = {Optics Express},
  volume = {19},
  number = {18},
  pages = {17189--17202},
  publisher = {Optica Publishing Group},
  issn = {1094-4087},
  doi = {10.1364/OE.19.017189},
  urldate = {2025-10-21},
  copyright = {\copyright{} 2011 OSA},
  langid = {english}
}

@article{ihle_stochastic_2003,
  title = {Stochastic Rotation Dynamics. {{I}}. {{Formalism}}, {{Galilean}} Invariance, and {{Green-Kubo}} Relations},
  author = {Ihle, T. and Kroll, D. M.},
  year = 2003,
  month = jun,
  journal = {Physical Review E},
  volume = {67},
  number = {6},
  pages = {066705},
  publisher = {American Physical Society},
  doi = {10.1103/PhysRevE.67.066705},
  urldate = {2025-01-17}
}

@article{irrgang_virial_2017,
  title = {Virial {{Coefficients}} and {{Equations}} of {{State}} for {{Hard Polyhedron Fluids}}},
  author = {Irrgang, M. Eric and Engel, Michael and Schultz, Andrew J. and Kofke, David A. and Glotzer, Sharon C.},
  year = 2017,
  month = oct,
  journal = {Langmuir},
  volume = {33},
  number = {42},
  pages = {11788--11796},
  publisher = {American Chemical Society},
  issn = {0743-7463},
  doi = {10.1021/acs.langmuir.7b02384},
  urldate = {2026-02-10}
}

@article{kammerer_dynamics_1997,
  title = {Dynamics of the Rotational Degrees of Freedom in a Supercooled Liquid of Diatomic Molecules},
  author = {K{\"a}mmerer, Stefan and Kob, Walter and Schilling, Rolf},
  year = 1997,
  month = nov,
  journal = {Physical Review E},
  volume = {56},
  number = {5},
  pages = {5450--5461},
  publisher = {American Physical Society},
  doi = {10.1103/PhysRevE.56.5450},
  urldate = {2025-11-13}
}

@incollection{kapral_multiparticle_2008,
  title = {Multiparticle {{Collision Dynamics}}: {{Simulation}} of {{Complex Systems}} on {{Mesoscales}}},
  shorttitle = {Multiparticle {{Collision Dynamics}}},
  booktitle = {Advances in {{Chemical Physics}}},
  author = {Kapral, Raymond},
  year = 2008,
  pages = {89--146},
  publisher = {John Wiley \& Sons, Ltd},
  doi = {10.1002/9780470371572.ch2},
  urldate = {2024-12-11},
  copyright = {Copyright \copyright{} 2008 John Wiley \& Sons, Inc.},
  isbn = {978-0-470-37157-2},
  langid = {english}
}

@article{kikuchi_transport_2003,
  title = {Transport Coefficients of a Mesoscopic Fluid Dynamics Model},
  author = {Kikuchi, N. and Pooley, C. M. and Ryder, J. F. and Yeomans, J. M.},
  year = 2003,
  month = sep,
  journal = {The Journal of Chemical Physics},
  volume = {119},
  number = {12},
  pages = {6388--6395},
  issn = {0021-9606},
  doi = {10.1063/1.1603721},
  urldate = {2026-04-14}
}

@article{kobayashi_self-assembly_2022,
  title = {Self-{{Assembly}} of {{Amphiphilic Cubes}} in {{Suspension}}},
  author = {Kobayashi, Yusei and Nikoubashman, Arash},
  year = 2022,
  month = aug,
  journal = {Langmuir},
  volume = {38},
  number = {34},
  pages = {10642--10648},
  publisher = {American Chemical Society},
  issn = {0743-7463},
  doi = {10.1021/acs.langmuir.2c01614},
  urldate = {2025-09-17}
}

@article{kobayashi_structure_2020,
  title = {Structure and Dynamics of Amphiphilic {{Janus}} Spheres and Spherocylinders under Shear},
  author = {Kobayashi, Yusei and Arai, Noriyoshi and Nikoubashman, Arash},
  year = 2020,
  journal = {Soft Matter},
  volume = {16},
  number = {2},
  pages = {476--486},
  publisher = {Royal Society of Chemistry},
  doi = {10.1039/C9SM01937E},
  urldate = {2025-09-16},
  langid = {english}
}

@article{kobayashi_structure_2020-1,
  title = {Structure and {{Shear Response}} of {{Janus Colloid}}--{{Polymer Mixtures}} in {{Solution}}},
  author = {Kobayashi, Yusei and Arai, Noriyoshi and Nikoubashman, Arash},
  year = 2020,
  month = dec,
  journal = {Langmuir},
  volume = {36},
  number = {47},
  pages = {14214--14223},
  publisher = {American Chemical Society},
  issn = {0743-7463},
  doi = {10.1021/acs.langmuir.0c02308},
  urldate = {2025-09-16}
}

@article{kundu_exploring_2025,
  title = {Exploring the Role of Hydrodynamic Interactions in Spherically Confined Drying Colloidal Suspensions},
  author = {Kundu, Mayukh and Kritika, Kritika and Wani, Yashraj M. and Nikoubashman, Arash and Howard, Michael P.},
  year = 2025,
  month = apr,
  journal = {The Journal of Chemical Physics},
  volume = {162},
  number = {15},
  pages = {154904},
  issn = {0021-9606},
  doi = {10.1063/5.0260883},
  urldate = {2025-09-16}
}

@article{linke_rotational_2018,
  title = {Rotational {{Diffusion Depends}} on {{Box Size}} in {{Molecular Dynamics Simulations}}},
  author = {Linke, Max and K{\"o}finger, J{\"u}rgen and Hummer, Gerhard},
  year = 2018,
  month = jun,
  journal = {The Journal of Physical Chemistry Letters},
  volume = {9},
  number = {11},
  pages = {2874--2878},
  publisher = {American Chemical Society},
  doi = {10.1021/acs.jpclett.8b01090},
  urldate = {2026-05-18}
}

@article{liu_dissipative_2015,
  title = {Dissipative {{Particle Dynamics}} ({{DPD}}): {{An Overview}} and {{Recent Developments}}},
  shorttitle = {Dissipative {{Particle Dynamics}} ({{DPD}})},
  author = {Liu, M. B. and Liu, G. R. and Zhou, L. W. and Chang, J. Z.},
  year = 2015,
  month = nov,
  journal = {Archives of Computational Methods in Engineering},
  volume = {22},
  number = {4},
  pages = {529--556},
  issn = {1886-1784},
  doi = {10.1007/s11831-014-9124-x},
  urldate = {2026-07-06},
  langid = {english}
}

@article{lobaskin_new_2004,
  title = {A New Model for Simulating Colloidal Dynamics},
  author = {Lobaskin, Vladimir and D{\"u}nweg, Burkhard},
  year = 2004,
  month = may,
  journal = {New Journal of Physics},
  volume = {6},
  number = {1},
  pages = {54},
  issn = {1367-2630},
  doi = {10.1088/1367-2630/6/1/054},
  urldate = {2025-09-11},
  langid = {english}
}

@article{malevanets_dynamics_2000,
  title = {Dynamics of Short Polymer Chains in Solution},
  author = {Malevanets, A. and Yeomans, J. M.},
  year = 2000,
  month = oct,
  journal = {Europhysics Letters},
  volume = {52},
  number = {2},
  pages = {231},
  publisher = {IOP Publishing},
  issn = {0295-5075},
  doi = {10.1209/epl/i2000-00428-0},
  urldate = {2025-11-17},
  langid = {english}
}

@article{malevanets_mesoscopic_1999,
  title = {Mesoscopic Model for Solvent Dynamics},
  author = {Malevanets, Anatoly and Kapral, Raymond},
  year = 1999,
  month = may,
  journal = {The Journal of Chemical Physics},
  volume = {110},
  number = {17},
  pages = {8605--8613},
  issn = {0021-9606},
  doi = {10.1063/1.478857},
  urldate = {2026-03-12}
}

@article{mauer_static_2017,
  title = {Static and Dynamic Light Scattering by Red Blood Cells: {{A}} Numerical Study},
  shorttitle = {Static and Dynamic Light Scattering by Red Blood Cells},
  author = {Mauer, Johannes and Peltom{\"a}ki, Matti and Poblete, Sim{\'o}n and Gompper, Gerhard and Fedosov, Dmitry A.},
  year = 2017,
  month = may,
  journal = {PLOS One},
  volume = {12},
  number = {5},
  pages = {e0176799},
  publisher = {Public Library of Science},
  issn = {1932-6203},
  doi = {10.1371/journal.pone.0176799},
  urldate = {2025-09-16},
  langid = {english}
}

@article{miller_symplectic_2002,
  title = {Symplectic Quaternion Scheme for Biophysical Molecular Dynamics},
  author = {Miller, III, T. F. and Eleftheriou, M. and Pattnaik, P. and Ndirango, A. and Newns, D. and Martyna, G. J.},
  year = 2002,
  month = may,
  journal = {The Journal of Chemical Physics},
  volume = {116},
  number = {20},
  pages = {8649--8659},
  issn = {0021-9606},
  doi = {10.1063/1.1473654},
  urldate = {2025-02-05}
}

@article{mo_method_1994,
  title = {A Method for Computing {{Stokes}} Flow Interactions among Spherical Objects and Its Application to Suspensions of Drops and Porous Particles},
  author = {Mo, Guobiao and Sangani, Ashok S.},
  year = 1994,
  month = may,
  journal = {Physics of Fluids},
  volume = {6},
  number = {5},
  pages = {1637--1652},
  issn = {1070-6631, 1089-7666},
  doi = {10.1063/1.868227},
  urldate = {2025-10-08},
  langid = {english}
}

@article{muller-plathe_reversing_1999,
  title = {Reversing the Perturbation in Nonequilibrium Molecular Dynamics: {{An}} Easy Way to Calculate the Shear Viscosity of Fluids},
  shorttitle = {Reversing the Perturbation in Nonequilibrium Molecular Dynamics},
  author = {{M{\"u}ller-Plathe}, Florian},
  year = 1999,
  month = may,
  journal = {Physical Review E},
  volume = {59},
  number = {5},
  pages = {4894--4898},
  publisher = {American Physical Society},
  doi = {10.1103/PhysRevE.59.4894},
  urldate = {2026-04-10}
}

@article{myung_weak_2018,
  title = {Weak {{Shape Anisotropy Leads}} to a {{Nonmonotonic Contribution}} to {{Crowding}}, {{Impacting Protein Dynamics}} under {{Physiologically Relevant Conditions}}},
  author = {Myung, Jin Suk and {Roosen-Runge}, Felix and Winkler, Roland G. and Gompper, Gerhard and Schurtenberger, Peter and Stradner, Anna},
  year = 2018,
  month = dec,
  journal = {The Journal of Physical Chemistry B},
  volume = {122},
  number = {51},
  pages = {12396--12402},
  publisher = {American Chemical Society},
  issn = {1520-6106},
  doi = {10.1021/acs.jpcb.8b07901},
  urldate = {2025-09-17}
}

@article{nguyen_rigid_2011,
  title = {Rigid Body Constraints Realized in Massively-Parallel Molecular Dynamics on Graphics Processing Units},
  author = {Nguyen, Trung Dac and Phillips, Carolyn L. and Anderson, Joshua A. and Glotzer, Sharon C.},
  year = 2011,
  month = nov,
  journal = {Computer Physics Communications},
  volume = {182},
  number = {11},
  pages = {2307--2313},
  issn = {0010-4655},
  doi = {10.1016/j.cpc.2011.06.005},
  urldate = {2025-10-14}
}

@misc{noauthor_mphowardlabazplugins_2026,
	howpublished = {https://github.com/mphowardlab/azplugins}
}

@article{noguchi_transport_2008,
  title = {Transport Coefficients of Off-Lattice Mesoscale-Hydrodynamics Simulation Techniques},
  author = {Noguchi, Hiroshi and Gompper, Gerhard},
  year = 2008,
  month = jul,
  journal = {Physical Review E},
  volume = {78},
  number = {1},
  pages = {016706},
  publisher = {American Physical Society},
  doi = {10.1103/PhysRevE.78.016706},
  urldate = {2025-06-12}
}

@article{okada_evaluation_2020,
  title = {Evaluation of the Translational and Rotational Diffusion Coefficients of a Cubic Particle (for the Application to {{Brownian}} Dynamics Simulations)},
  author = {Okada, Kazuya and Satoh, Akira},
  year = 2020,
  month = mar,
  journal = {Molecular Physics},
  volume = {118},
  number = {5},
  pages = {e1631498},
  publisher = {Taylor \& Francis},
  issn = {0026-8976},
  doi = {10.1080/00268976.2019.1631498},
  urldate = {2026-02-13}
}

@article{padding_hydrodynamic_2006,
  title = {Hydrodynamic Interactions and {{Brownian}} Forces in Colloidal Suspensions: {{Coarse-graining}} over Time and Length Scales},
  shorttitle = {Hydrodynamic Interactions and {{Brownian}} Forces in Colloidal Suspensions},
  author = {Padding, J. T. and Louis, A. A.},
  year = 2006,
  month = sep,
  journal = {Physical Review E},
  volume = {74},
  number = {3},
  pages = {031402},
  publisher = {American Physical Society},
  doi = {10.1103/PhysRevE.74.031402},
  urldate = {2025-09-19}
}

@article{padding_stick_2005,
  title = {Stick Boundary Conditions and Rotational Velocity Auto-Correlation Functions for Colloidal Particles in a Coarse-Grained Representation of the Solvent},
  author = {Padding, J T and Wysocki, A and L{\"o}wen, H and Louis, A A},
  year = 2005,
  month = oct,
  journal = {Journal of Physics: Condensed Matter},
  volume = {17},
  number = {45},
  pages = {S3393},
  issn = {0953-8984},
  doi = {10.1088/0953-8984/17/45/027},
  urldate = {2025-08-22},
  langid = {english}
}

@article{peng_multiparticle_2024,
  title = {Multiparticle Collision Dynamics Simulations of Hydrodynamic Interactions in Colloidal Suspensions: {{How}} Well Does the Discrete Particle Approach Do at Short Range?},
  shorttitle = {Multiparticle Collision Dynamics Simulations of Hydrodynamic Interactions in Colloidal Suspensions},
  author = {Peng, Ying-Shuo and Sinno, Talid},
  year = 2024,
  month = may,
  journal = {The Journal of Chemical Physics},
  volume = {160},
  number = {17},
  pages = {174121},
  issn = {0021-9606},
  doi = {10.1063/5.0197818},
  urldate = {2024-11-20}
}

@article{pettyjohn_effect_1948,
  title = {Effect of Particle Shape on Free Settling Rates of Isometric Particles},
  author = {Pettyjohn, E. A. and Christiansen, E. B.},
  year = 1948,
  journal = {Chemical Engineering Progress},
  volume = {44},
  pages = {157--172},
  issn = {0360-7275},
  urldate = {2026-05-21},
  langid = {english}
}

@article{poblete_hydrodynamics_2014,
  title = {Hydrodynamics of Discrete-Particle Models of Spherical Colloids: {{A}} Multiparticle Collision Dynamics Simulation Study},
  shorttitle = {Hydrodynamics of Discrete-Particle Models of Spherical Colloids},
  author = {Poblete, Sim{\'o}n and Wysocki, Adam and Gompper, Gerhard and Winkler, Roland G.},
  year = 2014,
  month = sep,
  journal = {Physical Review E},
  volume = {90},
  number = {3},
  pages = {033314},
  publisher = {American Physical Society},
  doi = {10.1103/PhysRevE.90.033314},
  urldate = {2024-11-20}
}

@article{ramasubramani_mean-field_2020,
  title = {A Mean-Field Approach to Simulating Anisotropic Particles},
  author = {Ramasubramani, Vyas and Vo, Thi and Anderson, Joshua A. and Glotzer, Sharon C.},
  year = 2020,
  month = aug,
  journal = {The Journal of Chemical Physics},
  volume = {153},
  number = {8},
  pages = {084106},
  issn = {0021-9606},
  doi = {10.1063/5.0019735},
  urldate = {2026-04-28}
}

@article{ripoll_dynamic_2005,
  title = {Dynamic Regimes of Fluids Simulated by Multiparticle-Collision Dynamics},
  author = {Ripoll, M. and Mussawisade, K. and Winkler, R. G. and Gompper, G.},
  year = 2005,
  month = jul,
  journal = {Physical Review E},
  volume = {72},
  number = {1},
  pages = {016701},
  publisher = {American Physical Society},
  doi = {10.1103/PhysRevE.72.016701},
  urldate = {2025-07-02}
}

@article{ripoll_low-reynolds-number_2004,
  title = {Low-{{Reynolds-number}} Hydrodynamics of Complex Fluids Bymulti-Particle-Collision Dynamics},
  author = {Ripoll, M. and Mussawisade, K. and Winkler, R. G. and Gompper, G.},
  year = 2004,
  month = oct,
  journal = {Europhysics Letters},
  volume = {68},
  number = {1},
  pages = {106},
  publisher = {IOP Publishing},
  issn = {0295-5075},
  doi = {10.1209/epl/i2003-10310-1},
  urldate = {2026-05-01},
  langid = {english}
}

@article{rusenargun_influence_2023,
  title = {Influence of Shape on Heteroaggregation of Model Microplastics: A Simulation Study},
  shorttitle = {Influence of Shape on Heteroaggregation of Model Microplastics},
  author = {Argun, B. Ru{\c s}en and Statt, Antonia},
  year = 2023,
  journal = {Soft Matter},
  volume = {19},
  number = {42},
  pages = {8081--8090},
  publisher = {Royal Society of Chemistry},
  doi = {10.1039/D3SM01014G},
  urldate = {2025-04-16},
  langid = {english}
}

@article{satterly_moments_1958,
  title = {The {{Moments}} of {{Inertia}} of Some {{Polyhedra}}},
  author = {Satterly, John},
  year = 1958,
  month = feb,
  journal = {The Mathematical Gazette},
  volume = {42},
  number = {339},
  pages = {11--13},
  issn = {0025-5572, 2056-6328},
  doi = {10.2307/3608345},
  urldate = {2026-08-07},
  langid = {english}
}

@article{statt_unexpected_2019,
  title = {Unexpected Secondary Flows in Reverse Nonequilibrium Shear Flow Simulations},
  author = {Statt, Antonia and Howard, Michael P. and Panagiotopoulos, Athanassios Z.},
  year = 2019,
  month = apr,
  journal = {Physical Review Fluids},
  volume = {4},
  number = {4},
  pages = {043905},
  publisher = {American Physical Society},
  doi = {10.1103/PhysRevFluids.4.043905},
  urldate = {2026-04-14}
}

@article{tenney_limitations_2010,
  title = {Limitations and Recommendations for the Calculation of Shear Viscosity Using Reverse Nonequilibrium Molecular Dynamics},
  author = {Tenney, Craig M. and Maginn, Edward J.},
  year = 2010,
  month = jan,
  journal = {The Journal of Chemical Physics},
  volume = {132},
  number = {1},
  pages = {014103},
  issn = {0021-9606},
  doi = {10.1063/1.3276454},
  urldate = {2026-04-10}
}

@article{tokuyama_dynamics_1994,
  title = {Dynamics of Hard-Sphere Suspensions},
  author = {Tokuyama, Michio and Oppenheim, Irwin},
  year = 1994,
  month = jul,
  journal = {Physical Review E},
  volume = {50},
  number = {1},
  pages = {R16-R19},
  publisher = {American Physical Society},
  doi = {10.1103/PhysRevE.50.R16},
  urldate = {2025-10-01}
}

@article{tuzel_transport_2003,
  title = {Transport Coefficients for Stochastic Rotation Dynamics in Three Dimensions},
  author = {T{\"u}zel, E. and Strauss, M. and Ihle, T. and Kroll, D. M.},
  year = 2003,
  month = sep,
  journal = {Physical Review E},
  volume = {68},
  number = {3},
  pages = {036701},
  publisher = {American Physical Society},
  doi = {10.1103/PhysRevE.68.036701},
  urldate = {2026-04-14}
}

@article{verberg_viscosity_1997,
  title = {Viscosity of Colloidal Suspensions},
  author = {Verberg, R. and {de Schepper}, I. M. and Cohen, E. G. D.},
  year = 1997,
  month = mar,
  journal = {Physical Review E},
  volume = {55},
  number = {3},
  pages = {3143--3158},
  publisher = {American Physical Society},
  doi = {10.1103/PhysRevE.55.3143},
  urldate = {2025-09-26}
}

@article{vogele_finite-size-corrected_2019,
  title = {Finite-{{Size-Corrected Rotational Diffusion Coefficients}} of {{Membrane Proteins}} and {{Carbon Nanotubes}} from {{Molecular Dynamics Simulations}}},
  author = {V{\"o}gele, Martin and K{\"o}finger, J{\"u}rgen and Hummer, Gerhard},
  year = 2019,
  month = jun,
  journal = {The Journal of Physical Chemistry B},
  volume = {123},
  number = {24},
  pages = {5099--5106},
  publisher = {American Chemical Society},
  issn = {1520-6106},
  doi = {10.1021/acs.jpcb.9b01656},
  urldate = {2026-05-18}
}

@misc{wales_global_2006,
  title = {Global Minima for the {{Thomson}} Problem},
  author = {Wales, David J. and Ulker, Sidika},
  year = 2006,
  month = dec,
  journal = {Global Minima for the Thomson Problem},
  urldate = {2026-04-16},
  howpublished = {https://www-wales.ch.cam.ac.uk/\textasciitilde wales/CCD/Thomson/table.html},
  langid = {english},
  note = {Accessed April 16, 2026}
}

@article{wales_structure_2006,
  title = {Structure and Dynamics of Spherical Crystals Characterized for the {{Thomson}} Problem},
  author = {Wales, David J. and Ulker, Sidika},
  year = 2006,
  month = dec,
  journal = {Physical Review B},
  volume = {74},
  number = {21},
  pages = {212101},
  publisher = {American Physical Society},
  doi = {10.1103/PhysRevB.74.212101},
  urldate = {2026-04-16}
}

@article{wang_short-time_2015-1,
  title = {Short-Time Transport Properties of Bidisperse Suspensions and Porous Media: {{A Stokesian}} Dynamics Study},
  shorttitle = {Short-Time Transport Properties of Bidisperse Suspensions and Porous Media},
  author = {Wang, Mu and Brady, John F.},
  year = 2015,
  month = mar,
  journal = {The Journal of Chemical Physics},
  volume = {142},
  number = {9},
  pages = {094901},
  issn = {0021-9606},
  doi = {10.1063/1.4913518},
  urldate = {2025-10-01}
}

@article{wani_diffusion_2022,
  title = {Diffusion and Sedimentation in Colloidal Suspensions Using Multiparticle Collision Dynamics with a Discrete Particle Model},
  author = {Wani, Yashraj M. and Kovakas, Penelope Grace and Nikoubashman, Arash and Howard, Michael P.},
  year = 2022,
  month = jan,
  journal = {The Journal of Chemical Physics},
  volume = {156},
  number = {2},
  pages = {024901},
  issn = {0021-9606},
  doi = {10.1063/5.0075002},
  urldate = {2024-11-20}
}

@article{wani_mesoscale_2024,
  title = {Mesoscale Simulations of Diffusion and Sedimentation in Shape-Anisotropic Nanoparticle Suspensions},
  author = {Wani, Yashraj M. and Kovakas, Penelope Grace and Nikoubashman, Arash and Howard, Michael P.},
  year = 2024,
  month = may,
  journal = {Soft Matter},
  volume = {20},
  number = {19},
  pages = {3942--3953},
  publisher = {The Royal Society of Chemistry},
  issn = {1744-6848},
  doi = {10.1039/D4SM00271G},
  urldate = {2024-11-20},
  langid = {english}
}

@article{weeks_role_1971,
  title = {Role of {{Repulsive Forces}} in {{Determining}} the {{Equilibrium Structure}} of {{Simple Liquids}}},
  author = {Weeks, John D. and Chandler, David and Andersen, Hans C.},
  year = 1971,
  month = jun,
  journal = {The Journal of Chemical Physics},
  volume = {54},
  number = {12},
  pages = {5237--5247},
  issn = {0021-9606},
  doi = {10.1063/1.1674820},
  urldate = {2026-06-23}
}

@article{whitmer_fluidsolid_2010,
  title = {Fluid--Solid Boundary Conditions for Multiparticle Collision Dynamics},
  author = {Whitmer, Jonathan K. and Luijten, Erik},
  year = 2010,
  month = feb,
  journal = {Journal of Physics: Condensed Matter},
  volume = {22},
  number = {10},
  pages = {104106},
  issn = {0953-8984},
  doi = {10.1088/0953-8984/22/10/104106},
  urldate = {2024-11-20},
  langid = {english}
}

@article{yang_effect_2015,
  title = {Effect of Angular Momentum Conservation on Hydrodynamic Simulations of Colloids},
  author = {Yang, Mingcheng and Theers, Mario and Hu, Jinglei and Gompper, Gerhard and Winkler, Roland G. and Ripoll, Marisol},
  year = 2015,
  month = jul,
  journal = {Physical Review E},
  volume = {92},
  number = {1},
  pages = {013301},
  publisher = {American Physical Society},
  doi = {10.1103/PhysRevE.92.013301},
  urldate = {2025-05-01}
}

@article{yeh_system-size_2004,
  title = {System-{{Size Dependence}} of {{Diffusion Coefficients}} and {{Viscosities}} from {{Molecular Dynamics Simulations}} with {{Periodic Boundary Conditions}}},
  author = {Yeh, In-Chul and Hummer, Gerhard},
  year = 2004,
  month = oct,
  journal = {The Journal of Physical Chemistry B},
  volume = {108},
  number = {40},
  pages = {15873--15879},
  publisher = {American Chemical Society},
  issn = {1520-6106},
  doi = {10.1021/jp0477147},
  urldate = {2025-10-06}
}

@article{yetkin_structure_2024,
  title = {Structure {{Formation}} in {{Supraparticles Composed}} of {{Spherical}} and {{Elongated Particles}}},
  author = {Yetkin, Melis and Wani, Yashraj M. and Kritika, Kritika and Howard, Michael P. and Kappl, Michael and Butt, Hans-J{\"u}rgen and Nikoubashman, Arash},
  year = 2024,
  month = jan,
  journal = {Langmuir},
  volume = {40},
  number = {1},
  pages = {1096--1108},
  publisher = {American Chemical Society},
  issn = {0743-7463},
  doi = {10.1021/acs.langmuir.3c03410},
  urldate = {2025-09-16}
}

\end{document}


\title{Supplementary material for ``Discretization strategies for colloidal particles in multiparticle collision dynamics simulations''}

\author{Michaela Bush}
\affiliation{Department of Chemical Engineering, Auburn University, Auburn, AL 36849, USA}

\author{Michael P. Howard}
\email{mphoward@auburn.edu}
\affiliation{Department of Chemical Engineering, Auburn University, Auburn, AL 36849, USA}

\maketitle

\begin{table}
\caption{Total number of excluded-volume sites for each regular polyhedron using outer or inner exclusion with $\sigma = 1.0\,\ell$ or $0.5\,\ell$.}
\label{tab:Npointpoly}
\begin{tabular}{lcccc}
& \multicolumn{2}{c}{outer exclusion} & \multicolumn{2}{c}{inner exclusion} \\
particle & $\sigma = 1.0\,\ell$ & $0.5\,\ell$ & $1.0\,\ell$ & $0.5\,\ell$ \\
\hline
cube & 386 & 1538 & 296 & 1352 \\
octahedron & 258 & 1026 & 198 & 902 \\
tetrahedron & 130 & 514 & 100 & 452
\end{tabular}
\end{table}

\section{Hydrodynamic Discretization}
\begin{figure}[!ht]
  \centering
  \includegraphics{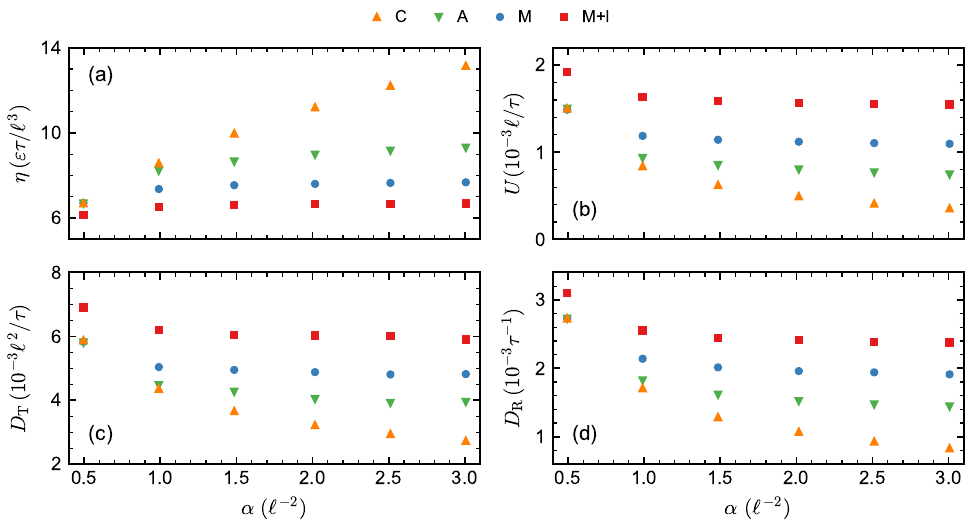}
  \caption{(a) Shear viscosity $\eta$, (b) sedimentation velocity $U$, (c) long-time translational self-diffusion coefficient $D_{\rm T}$, and (d) long-time rotational self-diffusion coefficient $D_{\rm R}$ for spheres with $d=3\,\ell$ when $\phi = 0.20$ using different surface-site densities $\alpha$ and strategies for selecting the surface-site mass $m_{\rm s}$.}
\end{figure}

\begin{figure}[!ht]
  \centering
  \includegraphics{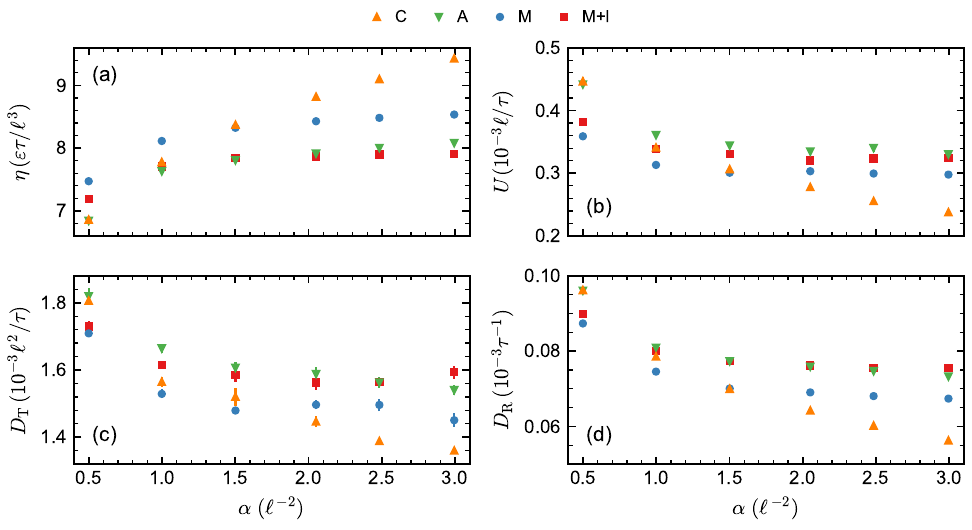}
  \caption{(a) Shear viscosity $\eta$, (b) sedimentation velocity $U$, (c) long-time translational self-diffusion coefficient $D_{\rm T}$, and (d) long-time rotational self-diffusion coefficient $D_{\rm R}$ for spheres with $d=9\,\ell$ when $\phi = 0.20$ using different surface-site densities $\alpha$ and strategies for selecting the surface-site mass $m_{\rm s}$.}
\end{figure}

\begin{figure}[!ht]
  \centering
  \includegraphics{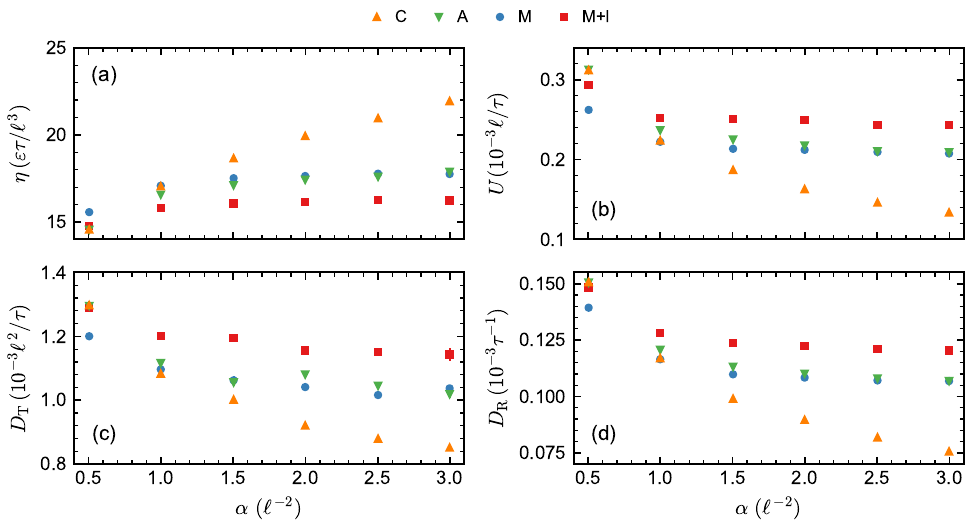}
  \caption{(a) Shear viscosity $\eta$, (b) sedimentation velocity $U$, (c) long-time translational self-diffusion coefficient $D_{\rm T}$, and (d) long-time rotational self-diffusion coefficient $D_{\rm R}$ for spheres with $d=6\,\ell$ when $\phi = 0.20$ using different surface-site densities $\alpha$ and strategies for selecting the surface-site mass $m_{\rm s}$. Here, the solvent density was $\rho_0 = 10\,m/\ell^3$ instead of $5\,m/\ell^{-3}$, giving $\eta_0 = 8.70\,\varepsilon \tau/\ell^3$, and the values of $m_s$ were adjusted to this density.}
\end{figure}

\begin{figure}[!ht]
  \centering
  \includegraphics{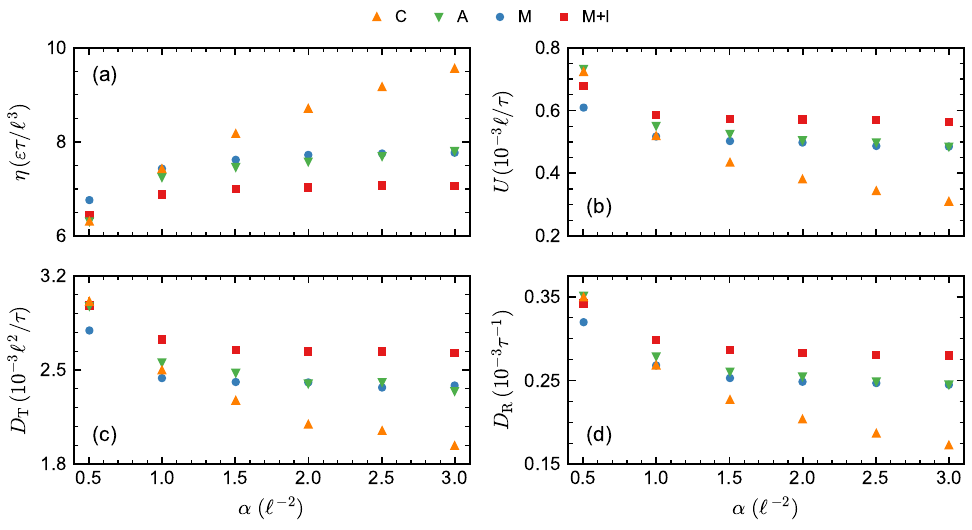}
  \caption{(a) Shear viscosity $\eta$, (b) sedimentation velocity $U$, (c) long-time translational self-diffusion coefficient $D_{\rm T}$, and (d) long-time rotational self-diffusion coefficient $D_{\rm R}$ for spheres with $d=6\,\ell$ when $\phi = 0.20$ using different surface-site densities $\alpha$ and strategies for selecting the surface-site mass $m_{\rm s}$. Here, the Andersen thermostat (AT) collision rule was used instead of the SRD collision rule, giving $\eta_0 = 3.71\,\varepsilon \tau/\ell^3$.\textsuperscript{67}}
\end{figure}

\begin{figure}[!ht]
  \centering
  \includegraphics{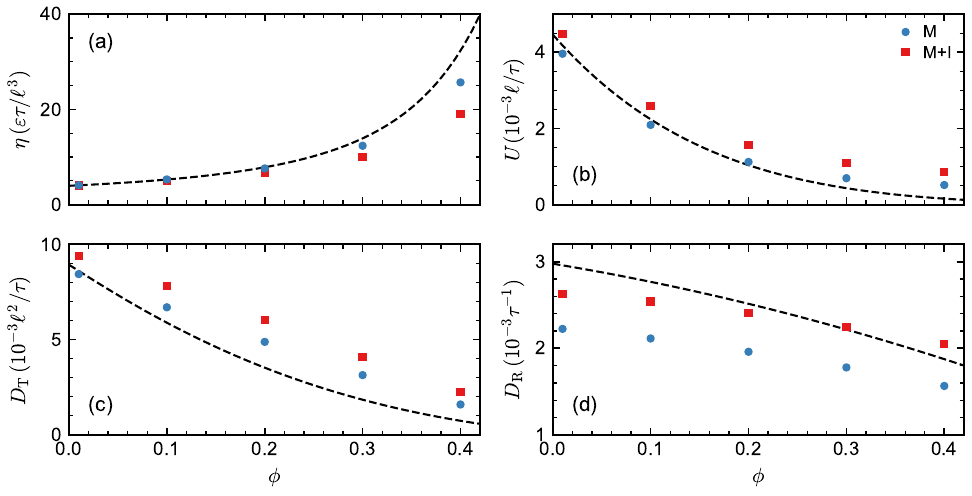}
  \caption{(a) Shear viscosity $\eta$, (b) sedimentation velocity $U$, (c) long-time translational self-diffusion coefficient $D_{\rm T}$, and (d) long-time rotational self-diffusion coefficient $D_{\rm R}$ for spheres with $d=3\,\ell$ spheres at varying volume fraction $\phi$ using $\alpha=2.0\,\ell^{-2}$ with the M and M+I strategies for selecting the surface-site mass $m_{\rm s}$. The dashed black lines are theoretical predictions of (a) Eq.~(6), (b) Eq.~(8), (c) Eq.~(9), and (d) Eq.~(11).}
\end{figure}

\begin{figure}[!ht]
  \centering
  \includegraphics{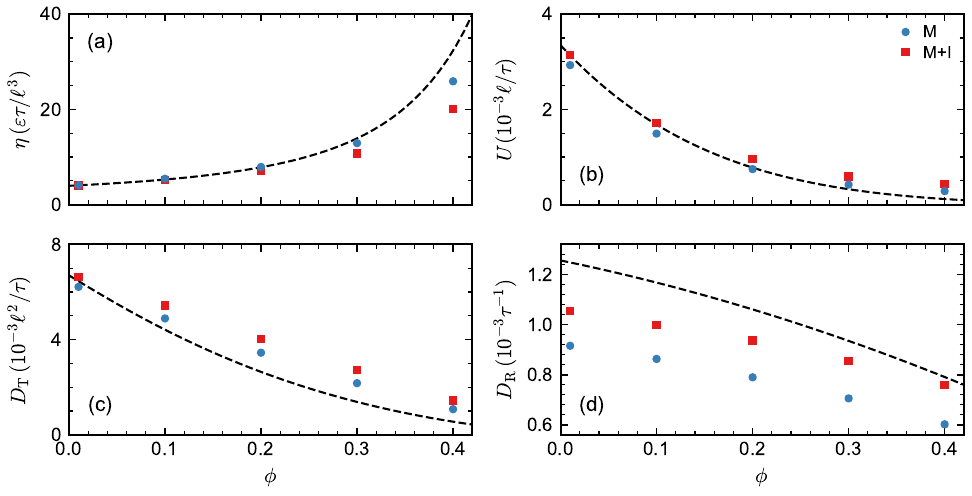}
  \caption{(a) Shear viscosity $\eta$, (b) sedimentation velocity $U$, (c) long-time translational self-diffusion coefficient $D_{\rm T}$, and (d) long-time rotational self-diffusion coefficient $D_{\rm R}$ for spheres with $d=4\,\ell$ spheres at varying volume fraction $\phi$ using $\alpha=2.0\,\ell^{-2}$ with the M and M+I strategies for selecting the surface-site mass $m_{\rm s}$. The dashed black lines are theoretical predictions of (a) Eq.~(6), (b) Eq.~(8), (c) Eq.~(9), and (d) Eq.~(11).}
\end{figure}

\begin{figure}[!ht]
  \centering
  \includegraphics{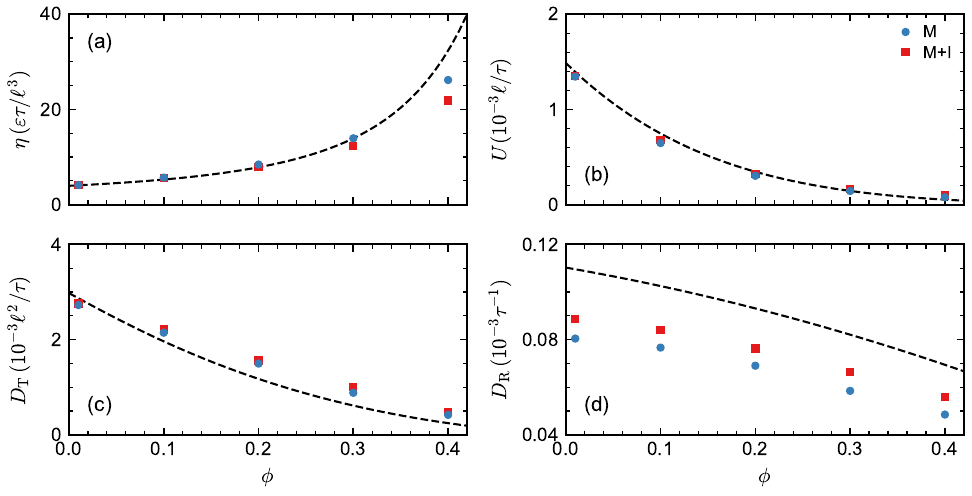}
  \caption{(a) Shear viscosity $\eta$, (b) sedimentation velocity $U$, (c) long-time translational self-diffusion coefficient $D_{\rm T}$, and (d) long-time rotational self-diffusion coefficient $D_{\rm R}$ for spheres with $d=9\,\ell$ spheres at varying volume fraction $\phi$ using $\alpha=2.0\,\ell^{-2}$ with the M and M+I strategies for selecting the surface-site mass $m_{\rm s}$. The dashed black lines are theoretical predictions of (a) Eq.~(6), (b) Eq.~(8), (c) Eq.~(9), and (d) Eq.~(11).}
\end{figure}

\begin{figure}[!ht]
  \centering
  \includegraphics{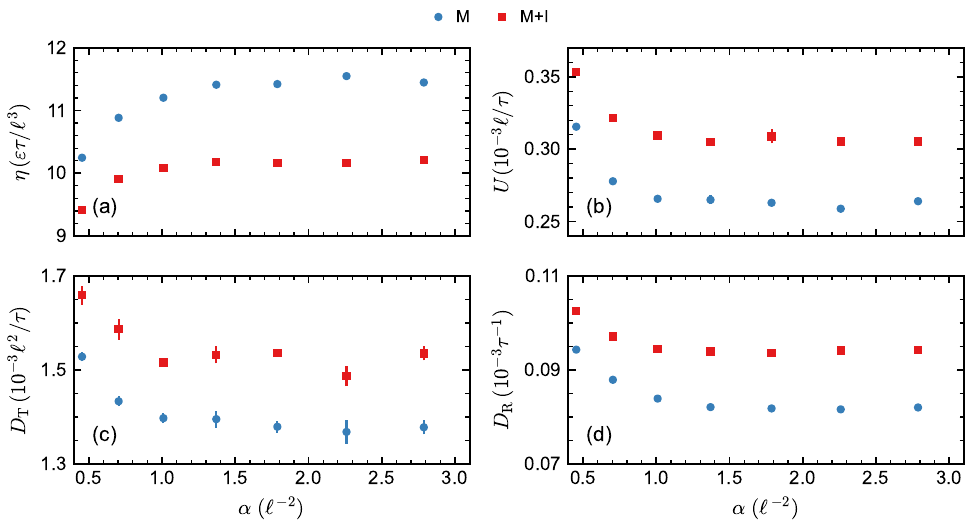}
  \caption{(a) Shear viscosity $\eta$, (b) sedimentation velocity $U$, (c) long-time translational self-diffusion coefficient $D_{\rm T}$, and (d) long-time rotational self-diffusion coefficient $D_{\rm R}$ for cubes when $\phi = 0.20$ using different surface-site densities $\alpha$ and strategies for selecting the surface-site mass $m_{\rm s}$.}
  \label{fig:6_cube}
\end{figure}

\clearpage
\section{Excluded-volume discretization}
\begin{figure}[!ht]
  \centering
  \includegraphics{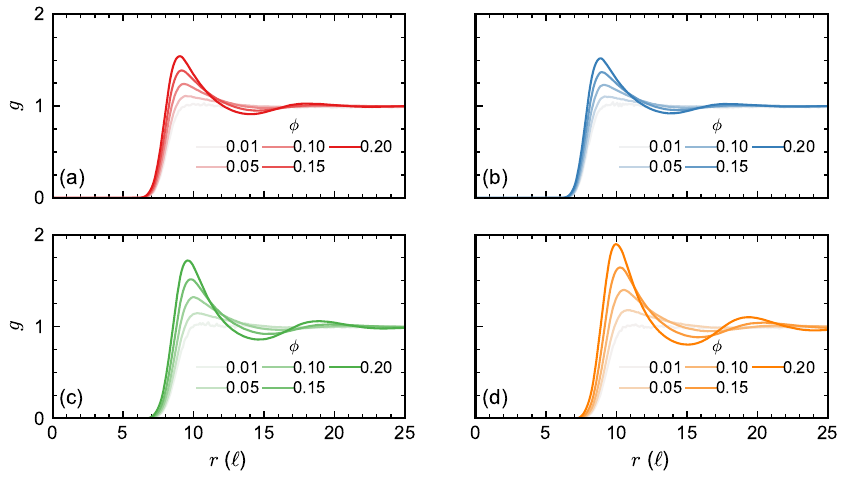}
  \caption{Radial distribution function $g$ for cubes as a function of distance $r$ using (a) inner exclusion with $\sigma=0.5\,\ell$, (b) inner exclusion with $\sigma=1.0\,\ell$, (c) outer exclusion with $\sigma=0.5\,\ell$, and (d) outer exclusion with $\sigma=1.0\,\ell$.}
  \label{fig:6_cube_rdf}
\end{figure}

\begin{figure}[!ht]
  \centering
  \includegraphics{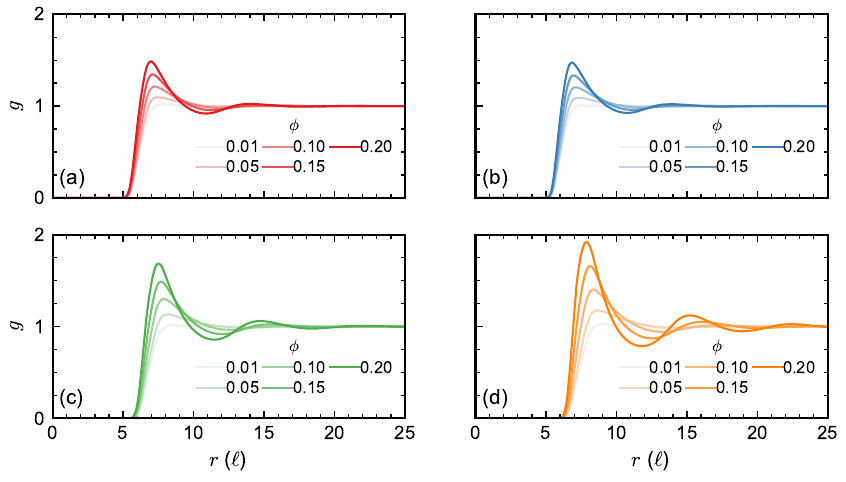}
  \caption{Radial distribution function $g$ for octahedra as a function of distance $r$ using (a) inner exclusion with $\sigma=0.5\,\ell$, (b) inner exclusion with $\sigma=1.0\,\ell$, (c) outer exclusion with $\sigma=0.5\,\ell$, and (d) outer exclusion with $\sigma=1.0\,\ell$.}
  \label{fig:6_octa_rdf}
\end{figure}

\begin{figure}[!ht]
  \centering
  \includegraphics{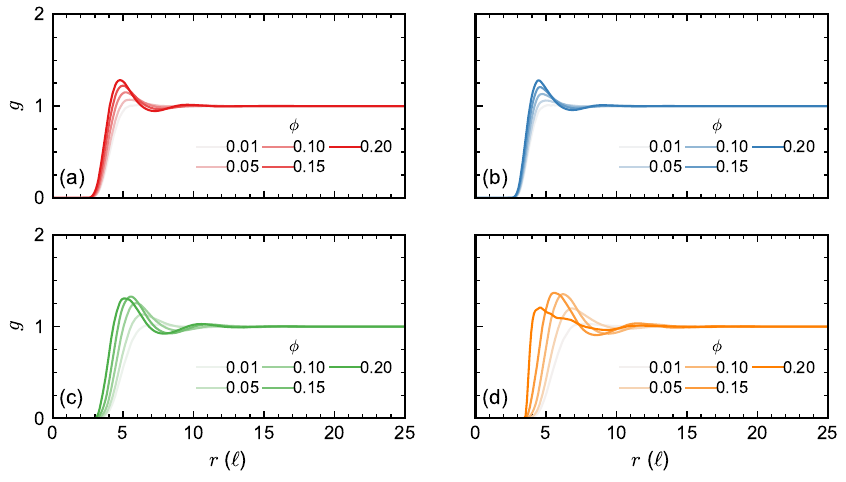}
  \caption{Radial distribution function $g$ for tetrahedra as a function of distance $r$ using (a) inner exclusion with $\sigma=0.5\,\ell$, (b) inner exclusion with $\sigma=1.0\,\ell$, (c) outer exclusion with $\sigma=0.5\,\ell$, and (d) outer exclusion with $\sigma=1.0\,\ell$. Note the emergence of a second peak in (d) when $\phi = 0.20$, which is associated with a phase transition.\textsuperscript{18,78,79}}
  \label{fig:6_tetra_rdf}
\end{figure}

\begin{figure}[!ht]
  \centering
  \includegraphics{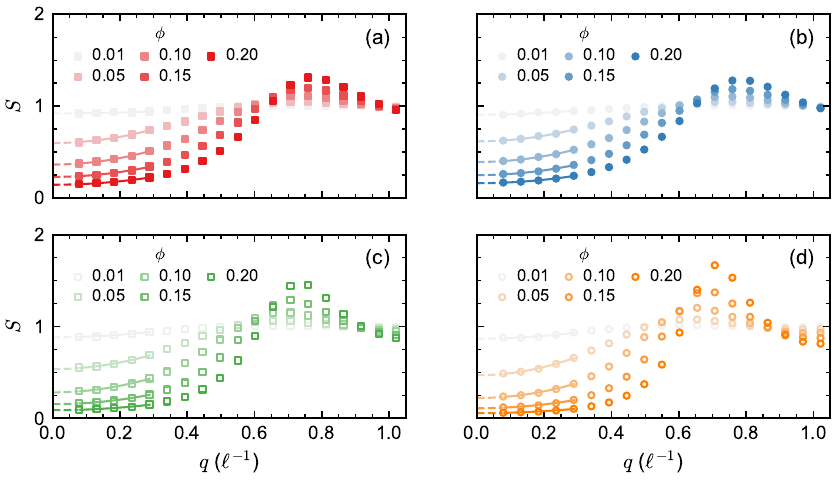}
  \caption{Static structure factor $S$ for cubes as a function of wavenumber $q$ using (a) inner exclusion with $\sigma=0.5\,\ell$, (b) inner exclusion with $\sigma=1.0\,\ell$, (c) outer exclusion with $\sigma=0.5\,\ell$, and (d) outer exclusion with $\sigma=1.0\,\ell$. The solid line is the fit described in the main text and the dashed lines are the extrapolation to $q=0$.}
  \label{fig:6_cube_sff}
\end{figure}

\begin{figure}[!ht]
  \centering
  \includegraphics{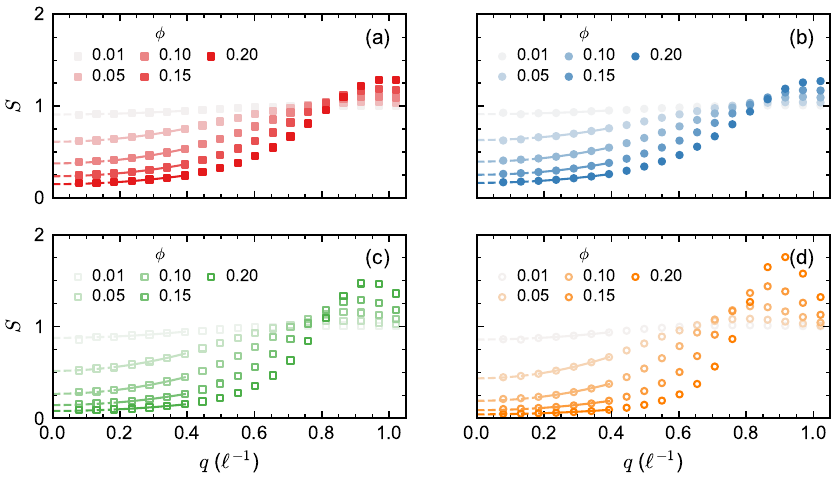}
  \caption{Static structure factor $S$ for octahedra as a function of wavenumber $q$ using (a) inner exclusion with $\sigma=0.5\,\ell$, (b) inner exclusion with $\sigma=1.0\,\ell$, (c) outer exclusion with $\sigma=0.5\,\ell$, and (d) outer exclusion with $\sigma=1.0\,\ell$. The solid line is the fit described in the main text and the dashed lines are the extrapolation to $q=0$.}
  \label{fig:6_octa_sff}
\end{figure}

\begin{figure}[!ht]
  \centering
  \includegraphics{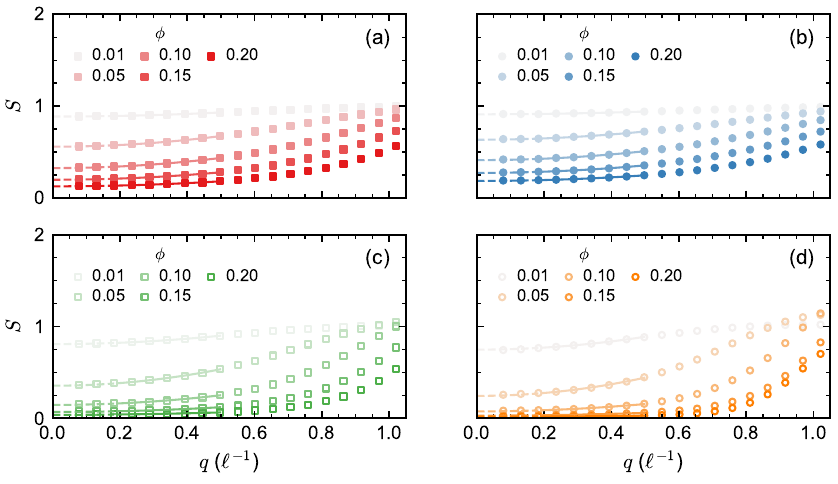}
  \caption{Structure factor $S$ for tetrahedra using (a) inner exclusion with $\sigma=0.5\,\ell$, (b) inner exclusion with $\sigma=1.0\,\ell$, (c) outer exclusion with $\sigma=0.5\,\ell$, and (d) outer exclusion with $\sigma=1.0\,\ell$. The solid line is the fit described in the main text and the dashed lines are the extrapolation to $q=0$.}
  \label{fig:6_tetra_sff}
\end{figure}

\begin{figure}[!ht]
  \centering
  \includegraphics{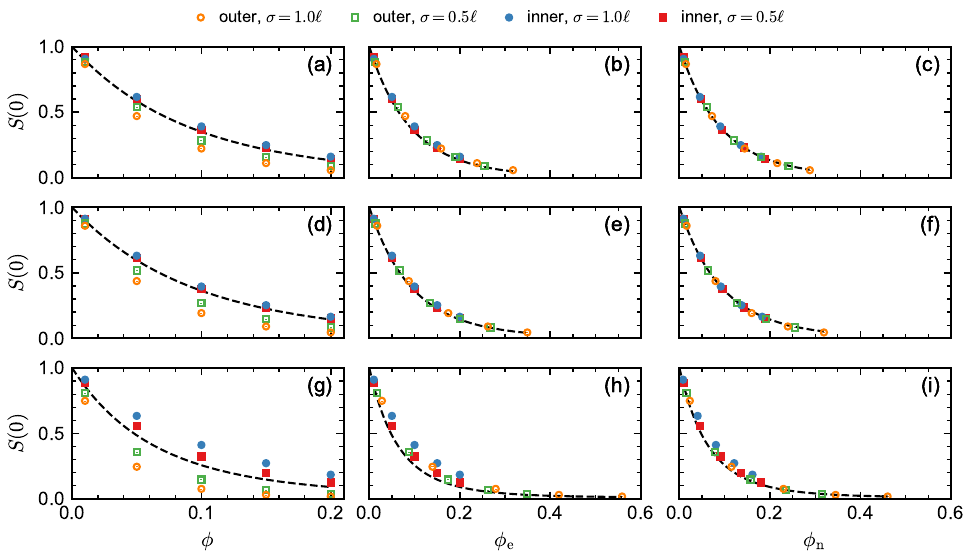}
  \caption{Static structure factor extrapolated to zero wavevector $S(0)$ as a function of volume fraction $\phi$ for (a--c) cubes, (d--f) octahedra, and (g--i) tetrahedra measured in isothermal--isochoric molecular dynamics simulations using outer and inner exclusion with $\sigma = 1.0\,\ell$ and $0.5\,\ell$ as a function of (a, d, g) nominal volume fraction $\phi$, (b, e, h) enclosing volume fraction $\phi_{\rm e}$, and (c, f, i) numerically estimated volume fraction $\phi_{\rm n}$. The dashed black lines are Eq.~(14) evaluated using the respective volume fractions.}
  \label{fig:s0}
\end{figure}

\begin{table}[ht]
\caption{Nominal volume $V_0$, enclosed volume $V_{\rm e}$, and numerically estimated volume $V_{\rm n}$ of the regular polyhedra for outer and inner exclusion with $\sigma = 1.0\,\ell$ and $0.5\,\ell$.}
\label{tab:volumes}
\begin{tabular}{cccccc}
shape & exclusion & $\sigma$ ($\ell$) & $V_0$ ($\ell^3$) & $V_{\rm e}$ ($\ell^3$) & $V_{\rm n}$ ($\ell^3$) \\
\hline
\multirow[c]{4}{*}{cube} & \multirow[c]{2}{*}{outer} & $1.0$ & 216.00 & 343.00 & 311.29 \\
 &  & $0.5$ & 216.00 & 274.62 & 260.57 \\
 & \multirow[c]{2}{*}{inner} & $1.0$ & 216.00 & 216.00 & 195.25 \\
 &  & $0.5$ & 216.00 & 216.00 & 204.70 \\
\multirow[c]{4}{*}{octahedron} & \multirow[c]{2}{*}{outer} & $1.0$ & 101.82 & 177.77 & 162.56 \\
 &  & $0.5$ & 101.82 & 136.29 & 129.74 \\
 & \multirow[c]{2}{*}{inner} & $1.0$ & 101.82 & 101.82 & 93.14 \\
 &  & $0.5$ & 101.82 & 101.82 & 96.92 \\
\multirow[c]{4}{*}{tetrahedron} & \multirow[c]{2}{*}{outer} & $1.0$ & 25.46 & 71.09 & 58.56 \\
 &  & $0.5$ & 25.46 & 44.44 & 40.07 \\
 & \multirow[c]{2}{*}{inner} & $1.0$ & 25.46 & 25.46 & 20.54 \\
 &  & $0.5$ & 25.46 & 25.46 & 23.04
\end{tabular}
\end{table}

\begin{figure*}
  \centering
  \includegraphics{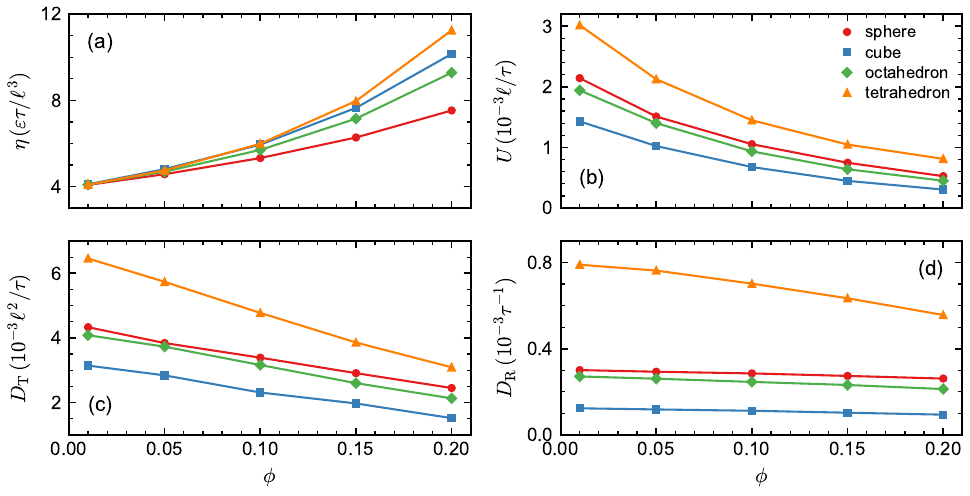}
  \caption{(a) Shear viscosity $\eta$, (b) sedimentation velocity $U$, (c) long-time translational self-diffusion coefficient $D_{\rm T}$, and (d) long-time rotational self-diffusion coefficient $D_{\rm R}$ for spheres with $d=6\,\ell$, cubes, octahedra, and tetrahedra as a function of the volume fraction $\phi$. The excluded-volume for the polyhedra was represented using inner exclusion with $\sigma = 1.0\,\ell$ for the cubes and octahedra and $\sigma = 0.5\,\ell$ for the tetrahedra.}
  \label{fig:chosen_phi}
\end{figure*}

\begin{figure}
  \centering
  \includegraphics{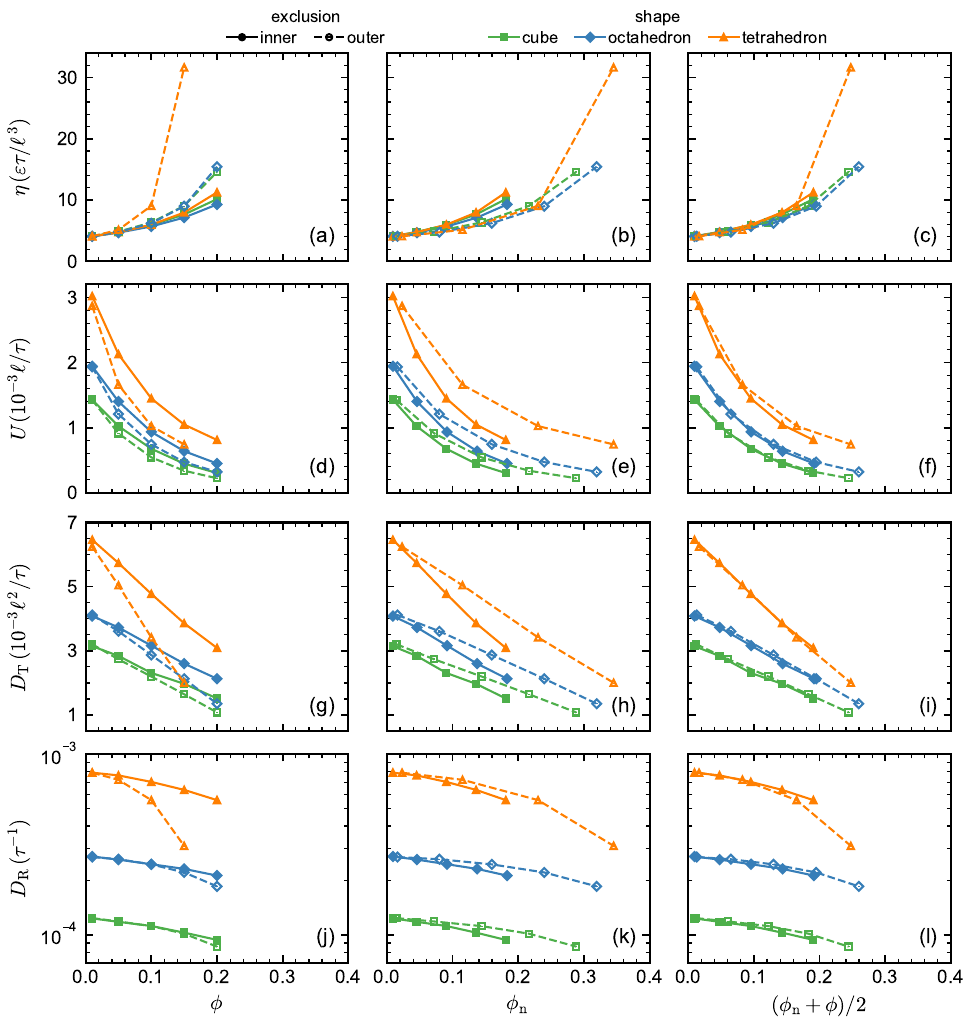}
  \caption{Transport properties (a--c) shear viscosity $\eta$, (d--f) sedimentation velocity $U$, (g--i) long-time translational diffusion coefficient $D_{\rm T}$, and (j-l) long-time rotational diffusion coefficient $D_{\rm R}$ for cubes, octahedra, and tetrahedra as a function of (a, d, g, j) nominal volume fraction $\phi$, (b, e, h, k) numerically estimated volume fraction $\phi_{\rm n}$, (c, f, i, l) and the average of $\phi$ and $\phi_{\rm n}$ . The solid lines and filled markers are inner exclusion with $\sigma=1.0\,\ell$ for the cubes and octahedra and $\sigma=0.5\,\ell$ for tetrahedra, while the dashed lines and unfilled markers are outer exclusion with $\sigma=1.0\,\ell$.}
  \label{fig:transport_exclusion}
\end{figure}